\documentclass[twocolumn, english, aps, showpacs, superscriptaddress]{revtex4-2}

\usepackage[T1]{fontenc}
\usepackage[latin9]{inputenc}
\usepackage{amsmath}
\usepackage{amssymb}
\usepackage{color}
\usepackage{times}
\usepackage{units}
\usepackage{amsmath}
\usepackage{amssymb}
\usepackage{graphicx}
\usepackage{wasysym}
\usepackage{soul} 
\usepackage{cancel}
\usepackage{babel}
\usepackage{hyperref}
\usepackage{dsfont}
\def\equationautorefname~#1\null{Equation (#1)\null}



\begin{document}
\title{Quantum interface for noble-gas spins based on spin-exchange collisions}
\author{Or Katz}
\thanks{These authors contributed equally to this work.}
\affiliation{Department of Physics of Complex Systems, Weizmann Institute of Science,
Rehovot 76100, Israel}
\affiliation{Rafael Ltd, IL-31021 Haifa, Israel}
\author{Roy Shaham}
\thanks{These authors contributed equally to this work.}
\affiliation{Department of Physics of Complex Systems, Weizmann Institute of Science,
Rehovot 76100, Israel}
\affiliation{Rafael Ltd, IL-31021 Haifa, Israel}
\author{Ofer Firstenberg}
\affiliation{Department of Physics of Complex Systems, Weizmann Institute of Science,
Rehovot 76100, Israel}

\begin{abstract}
An ensemble of noble-gas nuclear spins is a unique quantum system that could maintain coherence for many hours at room temperature and above, owing to exceptional isolation from the environment.
This isolation, however, is a mixed blessing, limiting the ability of these ensembles to interface with other quantum systems coherently.
Here we show that spin-exchange collisions with alkali-metal atoms render a quantum interface for noble-gas spins without impeding their long coherence times.
We formulate the many-body theory of the hybrid system and reveal a collective mechanism that strongly couples the macroscopic quantum states of the two spin ensembles.
Despite their stochastic and random nature, weak collisions enable entanglement and reversible exchange of nonclassical excitations in an efficient, controllable, and deterministic process.
With recent experiments now entering the strong-coupling regime, this interface paves the way towards realizing hour-long quantum memories and entanglement at room-temperature.

\end{abstract}

\maketitle

Macroscopic systems exhibiting quantum behavior at or above room temperature are of great scientific interest.
One such prominent system is a hot vapor of alkali-metal atoms enclosed in a vacuum cell.
The collective spin state of these ensembles, consisting
of as many as $10^{14}$ atoms, has been used to demonstrate quantum spin squeezing, storage and control of single excitation quanta, and entanglement at room temperature \cite{Julsgaard2001PolzikEntanglement,Polzik2010ReviewRMP,Sherson2006PolzikTeleportationDemo,Phillips2001LukinStorageInVapor,Fleischhauer2000LukinEITPolaritons,Eisaman2005LukinSinglePhoton,Hosseini2011BuchlerMemory,Finkelstein2018FirstenbergFlame,Kaczmarek2018NunnORCA,Ripka2018RydbergPfau,Happer2010book,Budker2013OpticalMagnetometryII,Balabas2010minutecoating,Katz2015SERFHybridization,Vasilakis2015PolzikBackactionEvation,Bao2020XiaoSpinSqueezing,ThomasPolzik2021AlkaliMechanics,Kong2018MitchellAlkaliSEEntanglement}.
Despite the rapid thermal motion and atomic collisions, the coherence time of the collective spin in these studies reaches milliseconds and beyond.
In some settings, it is unaffected by frequent spin-exchange collisions \cite{Kong2018MitchellAlkaliSEEntanglement,Kominis2008squeezingTheory,Dellis2014SESpinNoisePRA} and essentially limited by the electron spins' coupling to their surroundings.

Odd isotopes of noble gases, such as $^{3}$He, possess a nonzero nuclear spin.
This spin is optically-inaccessible and well protected from the external environment by the enclosing, full, electronic shells and is therefore extremely long-lived.
Noble-gas spin ensembles have demonstrated lifetimes $T_{1}$ exceeding hundreds of hours and coherence times $T_{2}^{*}$ of 100 hours at or above room temperature \cite{Chen2011He3NIST,Gemmel2010UltraSensitiveMagnetometer,Walker2017He3review,Heil2013T2100h}.
It is to be expected that the collective nuclear spin in these ensembles, similarly to the collective electronic spin in alkali vapor, can be brought to the quantum limit and utilized for quantum optics and sensing applications.
Several theory works have proposed to use collisions with optically-excited metastable $^{3}$He atoms for generating and reading-out nonclassical states of ground-level $^{3}$He spins, relying on  collisional-exchange of electronic configurations that lead to adiabatic transfer of the quantum state \cite{Dantan2005SinatraPinardMEOPstorage,Reinaudi2007SinatraPinardMEOPentanglement,SerafinSinatra2020squeezingPRL,SerafinSinatra2020squeezingPRA}.

This paper studies a different mechanism forming a quantum interface to noble-gas spins.
The mechanism relies on the stochastic accumulation of weak spin-exchange collisions, directly coupling between the collective spins of polarized atomic gases and thus not limited to adiabatic transfer rates.
Recently, we have experimentally demonstrated the coherent coupling between alkali-vapor spins and noble-gas spins, in the regime where the coherent exchange rate dominates the spin relaxation rate \cite{Shaham2021StrongCoupling}.
Additionally, we have demonstrated
a bi-directional interface between light and noble-gas spins, using the optically-accessible alkali spins as mediators \cite{Katz2021NobleSpectroscopy}.
These experiments establish the coherent nature of the collective coupling for classical (coherent) states.
If the coupling is indeed quantum, \textit{i.e.}, if it transfers non-classical correlations between the spin ensembles, then it can be used for quantum information applications, such as quantum memories and remote entanglement with long-lived noble-gas spins \cite{AlkaliNobleEntanglementKatz2020PRL,noblegasStoragePRA2020arxiv,noblegasStoragePRL2020arxiv}.

Here we provide a theoretical description of the emergent collective coupling between the spins of polarized alkali and noble-gas atoms.
We analyze the effect of weak spin-exchange collisions using a many-body formalism and find that they can efficiently couple the collective quantum excitations of the two ensembles, with negligible quantum noise added due to the collisions' stochasticity.
We derive the strong-coupling conditions, where the quantum-coherent exchange dominates over the relaxation.
We use numerical many-body simulations and a detailed analytical model to study the controllable periodic exchange of non-classical states between the spin ensembles, further attesting to the quantum nature of the interface.
Finally, we outline practical experimental conditions for the efficient exchange of non-classical states between the alkali and noble-gas ensembles.

\begin{figure*}[t]
\centering{}\includegraphics[width=16cm]{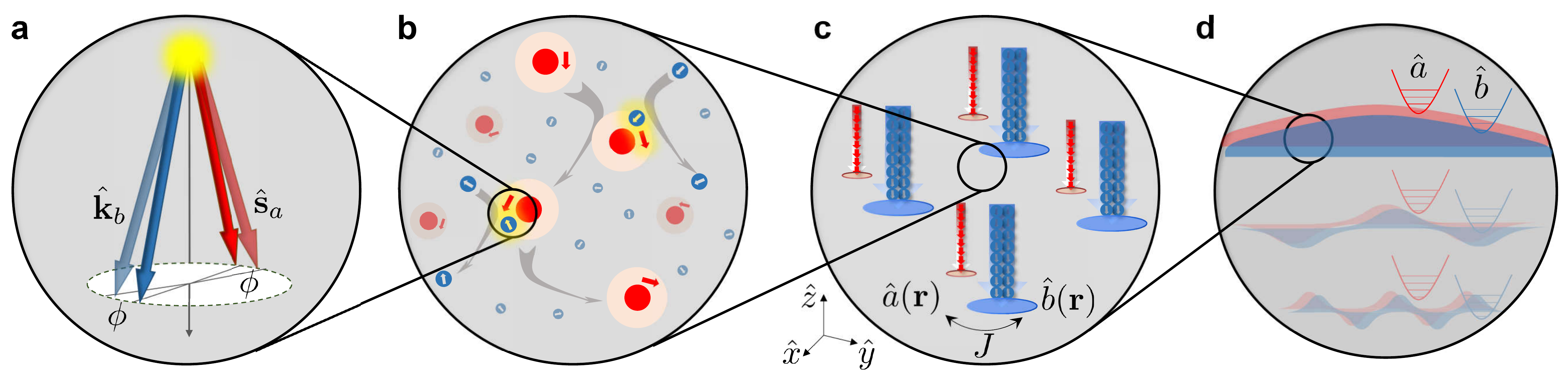}\caption{\textbf{Quantum interface for noble-gas spins via spin-exchange collisions.}
\textbf{a}, Coherent interaction during a collision between alkali-metal electronic spin $\boldsymbol{\hat{\textbf{s}}}_{a}$ (red) and noble-gas nuclear spin $\boldsymbol{\hat{\textbf{k}}}_{b}$ (blue).
The two spins mutually precess and acquire an angle $\phi\ll1$ while conserving the total spin, where $\phi$ is random and depends on the collision kinematics.
\textbf{b}, Stochastic sequence of collisions. Spin exchange occurs over a few picoseconds when the valence electron's wave-function (pink) overlaps with the noble-gas nucleus (blue).
\textbf{ c}, For polarized ensembles, multiple collisions between different atoms accumulate to a coherent dynamics of bosonic collective-spin excitations, described by local quantum operators $\hat{a}(\mathbf{r})$ (alkali) and $\hat{b}(\mathbf{r})$ (noble gas) and a coupling rate $J$.
Incoherent spin dynamics, which enables initialization via spin-exchange optical-pumping, play a minor role for $\phi\lll1$.
\textbf{d}, Diffusion of the gaseous atoms and the boundary conditions in the cell define nonlocal spin modes.
The lowest-order spatial modes $\hat{a},\hat{b}$ govern the coherent evolution of the collective quantum spins (see Methods, \cite{Shaham2020Diffusion}).\label{fig:illustration}}
\end{figure*}

\textbf{System.}
Noble-gas spins can be accessed via spin-exchange collisions with alkali vapor atoms.
Consider a gaseous mixture of $N_{\mathrm{b}}$ noble-gas atoms with nuclear spin-1/2 and $N_{\mathrm{a}}\ll N_{\mathrm{b}}$
alkali-metal atoms, all enclosed in a heated spherical cell of volume $V$ and undergoing frequent collisions.
A collision between noble-gas atom $b$ and alkali atom $a$ is governed by the Fermi-contact interaction and described by the evolution operator $\exp(-i\phi\boldsymbol{\hat{\textbf{k}}}_{b}\cdot\boldsymbol{\hat{\textbf{s}}}_{a})$.
$\boldsymbol{\hat{\textbf{k}}}_{b}$ is the noble-gas nuclear spin operator and $\boldsymbol{\hat{\textbf{s}}}_{a}$ is the alkali electronic spin operator of the colliding atoms labeled $a,b$ respectively.
$\phi$ is the mutual precession angle; see Figs.~\ref{fig:illustration}a and \ref{fig:illustration}b \cite{Walker1997SEOPReview,AppeltHapper1998SEOPtheoryPRA,Romalis2014CommentGradientsSphere}.
While $\phi$ varies between collisions depending on the atoms' random trajectories, its value is always positive \cite{Walter1998HapperWalkerPhiTrajectory,Schaefer1989WalkerKappa0}.
This is an important property of the isotropic Fermi-contact interaction, leading to a nonzero mean precession $\langle\phi\rangle$ during collisions.

Between collisions, the electron and nuclear spins of the \emph{alkali} atoms are altered by their strong hyperfine coupling.
Consequently, the slow dynamics of alkali atoms having nuclear spin $I>0$ is determined by the operator sum $\boldsymbol{\hat{\textbf{f}}}_{a}=\boldsymbol{\hat{\textbf{s}}}_{a}+\boldsymbol{\hat{\textbf{i}}}_{a}$, where $\boldsymbol{\hat{\textbf{i}}}_{a}$ is the nuclear spin operator of alkali atom $a$.
We shall focus on the conditions of high alkali atomic density and small Zeeman splitting.
Under these conditions, the alkali Zeeman states are populated with a spin-temperature distribution, and $\boldsymbol{\hat{\textbf{f}}}_{a}=q\boldsymbol{\hat{\textbf{s}}}_{a}$, where $q$ is known as the Larmor slowing-down factor [$2I+1<q<2I^2+I+1$ depending on the degree of polarization, see Eq.~(\ref{eq:slowdown_factor})] \cite{AppeltHapper1998SEOPtheoryPRA,Vasilakis2011RomalisBackactionEvation}.

The accepted formalism for describing the dynamics of the spin ensembles employs the mean-field Bloch equations \cite{Happer2010book}
\begin{equation}
\begin{aligned}\partial_{t}\langle\boldsymbol{\hat{\textbf{f}}}\rangle & =n_{\mathrm{b}}\zeta\langle\boldsymbol{\hat{\textbf{k}}}\rangle\times\langle\boldsymbol{\hat{\textbf{f}}}\rangle+n_{\mathrm{b}}k_{\text{se}}\bigl(q\langle\boldsymbol{\hat{\textbf{k}}}\rangle-\langle\boldsymbol{\hat{\textbf{f}}}\rangle\bigr),\\
\partial_{t}\langle\boldsymbol{\hat{\textbf{k}}}\rangle & =n_{\mathrm{a}}\zeta\langle\boldsymbol{\hat{\textbf{f}}}\rangle\times\langle\boldsymbol{\hat{\textbf{k}}}\rangle+n_{\mathrm{a}}k_{\text{se}}\bigl(\langle\boldsymbol{\hat{\textbf{f}}}\rangle/q-\langle\boldsymbol{\hat{\textbf{k}}}\rangle\bigr).
\end{aligned}
\label{eq:Bloch-dynamics}
\end{equation}
Here $n_{\mathrm{a}}=N_{\mathrm{a}}/V$ and $n_{\mathrm{b}}=N_{\mathrm{b}}/V$ are the atomic densities, $\zeta=\langle\sigma v\phi\rangle/q$ denotes the mean-field interaction strength, where $\sigma$ is the spin-exchange cross-section, $v$ is the relative thermal velocity, and $k_{\text{se}}\equiv\frac{1}{4}\bigl\langle v\sigma\phi^{2}\bigr\rangle$ is known as the binary spin-exchange rate coefficient \cite{Happer2010book}.
The first term in Eqs.~(\ref{eq:Bloch-dynamics}) describes the mutual average precession of the two mean spins.
The second term represents an incoherent transfer of spin polarization from one species to another (conserving the total spin) \cite{AppeltHapper1998SEOPtheoryPRA}; it is commonly utilized for initially polarizing the noble-gas spins via so-called spin-exchange optical pumping (SEOP) \cite{Walker1997SEOPReview}.
The mean-field Eqs.~(\ref{eq:Bloch-dynamics}) implicitly assume that any quantum-correlation developed between different atoms due to collisions is rapidly lost.
Therefore, this model is insufficient for describing the dynamics of nonclassical, \emph{i.e.} quantum, spin states.

\textbf{Dynamics of collective spin states.}
We describe the macroscopic quantum states of the spin ensembles using collective spin operators \cite{Polzik2010ReviewRMP}.
Each collective operator is a symmetric superposition of the spins in the cell, where $\boldsymbol{\hat{\textbf{F}}}=\sum_{a=1}^{N_{\text{a}}}\boldsymbol{\hat{\textbf{f}}}_{a}$ is the collective alkali spin operator, and $\boldsymbol{\hat{\textbf{K}}}=\sum_{b=1}^{N_{\text{b}}}\boldsymbol{\hat{\textbf{k}}}_{b}$ is the collective noble-gas spin operator.
Here we focus on the particular regime of highly-polarized ensembles, with most of the spins pointing downward ($-\hat{z}$) \cite{Julsgaard2001PolzikEntanglement,Polzik2010ReviewRMP,Sherson2006PolzikTeleportationDemo}.
In this regime, the operators $\hat{\mathrm{F}}_{z}$ and $\hat{\mathrm{K}}_{z}$ can be approximated by their classical expectation values $\mathrm{F}_{z}=-p_{\mathrm{a}}N_{\mathrm{a}}q/2$ and $\mathrm{K}_{z}=-p_{\mathrm{b}}N_{\mathrm{b}}/2$, where $p_{\mathrm{a}},p_{\mathrm{b}}\le1$ are the degrees of spin polarization ($p_{\mathrm{a}}=p_{\mathrm{b}}=1$ for the ideal preparation of fully-polarized ensembles).
The quantum state of the collective spins is then fully captured by the ladder operators $\hat{\mathrm{F}}^{\pm}=\hat{\mathrm{F}}_{x}\pm i\hat{\mathrm{F}}_{y}$ and $\hat{\mathrm{K}}^{\pm}=\hat{\mathrm{K}}_{x}\pm i\hat{\mathrm{K}}_{y}$.
Pictorially, for polarized ensembles, these operators describe a small tilt of the macroscopic spin vector, accompanied by spin uncertainty that scales as $\sqrt{N_{\text{a}}}$ for the alkali ensemble and $\sqrt{N_{\text{b}}}$ for the noble gas.

\begin{figure*}
\centering{}\includegraphics[width=\textwidth]{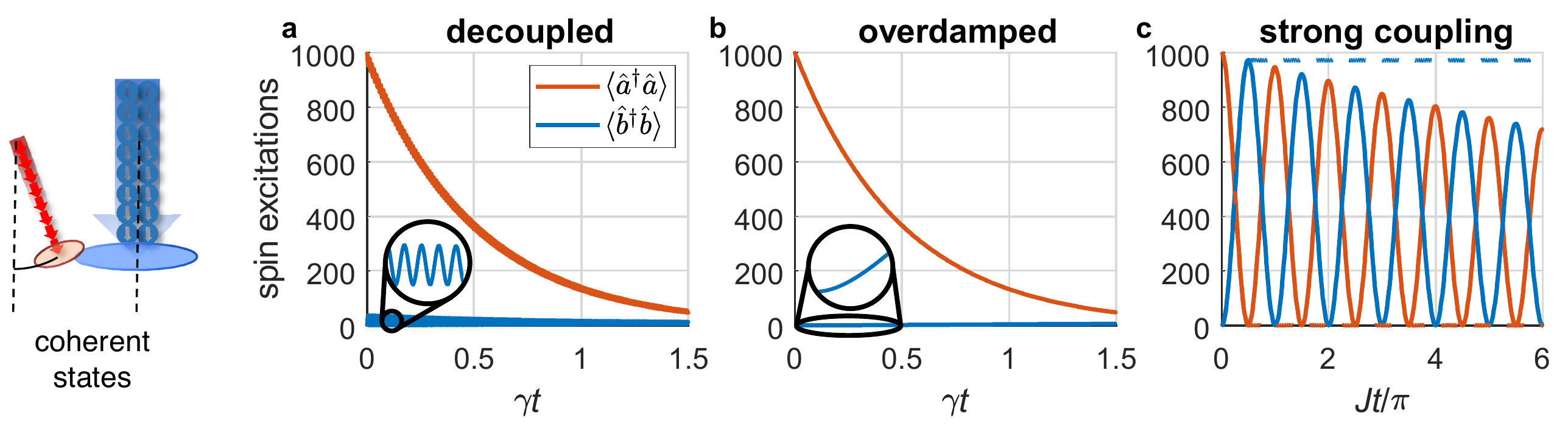}\caption{\textbf{Evolution of spin excitations of interacting alkali and noble-gas ensembles in different coupling regimes.}
Both ensembles are initialized in coherent spin states, with the alkali spins initially having $\langle\hat{a}^{\dagger}\hat{a}\rangle=1000$ excitations and the noble-gas spins in the vacuum state $\langle\hat{b}^{\dagger}\hat{b}\rangle=0$.
\textbf{a}, Decoupled modes.
At large detunings $\Delta=10J=600\gamma$, the spin ensembles are decoupled, and the alkali excitations decay at a rate $\gamma$.
The noble-gas spins exhibit negligible exchange with the alkali (inset).
\textbf{b}, Overdamped coupling.
At $\Delta=0$, the alkali and noble-gas spins hybridize and decay, here at a rate $\gamma=10J$, exhibiting partial transfer of the excitations (inset).
\textbf{c}, Strong coupling.
The periodic exchange for $J=57\gamma$ at $\Delta=0$ allows for coherent transfer of spin excitations.
Application of a large magnetic field at $t=\pi/(2J)$ decouples the spin ensembles and 'stores' the excitations in the noble-gas spin state (dashed line).
\label{fig:coupling modes} }
\end{figure*}

To see how quantum spin excitations are associated with these operators, we apply the Holstein-Primakoff transformation \cite{Polzik2010ReviewRMP}, which describes the collective states in terms of excitations of bosonic fields.
We define the annihilation operators of the two ensembles as $\hat{a}=\hat{\text{F}}^{-}/\sqrt{2|\text{F}_{z}|}$ and $\hat{b}=\hat{\text{K}}^{-}/\sqrt{2|\text{K}_{z}|}$.
The state $\left|0\right\rangle _{\mathrm{a}}\left|0\right\rangle _{\mathrm{b}}$, with zero spins pointing upwards, is identified as the vacuum, and the creation operators $\hat{a}^{\dagger}$ and $\hat{b}^{\dagger}$ flip upwards one alkali or noble-gas spin.

We model the quantum dynamics of the collective spins by following the steps illustrated in Fig.~\ref{fig:illustration}.
We adopt a microscopic picture of a stochastic sequence of random collisions and keep track of the correlations developed between different atoms during collisions (see Methods).
With the addition of a magnetic field $B\hat{z}$ and decoherence for the alkali spins at a rate $\gamma$, we find that the dynamics of the coupled quantum systems is well approximated by 
\begin{equation}
\begin{aligned}
\partial_{t}\hat{a} & =-iJ\hat{b}+(i\Delta-\gamma)\hat{a}+\hat{F}_{\mathrm{a}}\\
\partial_{t}\hat{b} & =-iJ\hat{a}.
\end{aligned}
\label{eq:coherent-dynamics}
\end{equation}
These equations are written in a rotating frame of the noble-gas spins, where $\Delta=(g_{\text{a}}-g_{\text{b}})B+\Delta_{\text{c}}$ denotes the mismatch of precession frequencies of the two polarized gases, with $g_{\text{a}}$ and $g_{\text{b}}$ being the gyromagnetic ratios of the alkali and noble-gas spins.
At zero magnetic field, the detuning is biased by the mean collisional shift $\Delta_{\text{c}}=\zeta(p_{\mathrm{a}}n_{\mathrm{a}}q-p_{\mathrm{b}}n_{\mathrm{b}})/2$ due to the difference in the effective magnetic fields induced by one species on the other \cite{Schaefer1989WalkerKappa0}.
$\hat{F}_{\mathrm{a}}$ is a quantum noise operator \cite{GardinerZoller2004quantumNoise}.
The decay of the noble-gas spins is omitted here, as we are interested in time scales much shorter than their (hours-long) decoherence time. 

A key result of our formalism is the identification of the bi-directional coupling rate $J=(\zeta/2)\sqrt{q p_{\mathrm{a}} p_{\mathrm{b}} n_{\mathrm{a}} n_{\mathrm{b}}}$, which represents the frequency at which the states of the two ensembles are exchanged.
Importantly, $J$ is proportional to the square-root of the atomic densities, which implies that the coupling is collective and benefits from collective enhancement.
For the special case of coherent spin states, one can reduce the quantum model [Eqs.~(\ref{eq:coherent-dynamics})] to the mean-field model [Eqs.~(\ref{eq:Bloch-dynamics})], by employing the transformations $\langle\hat{a}\rangle=\sqrt{N_{\mathrm{a}}/(q p_{\mathrm{a}} )}\langle\hat{\mathrm{f}}^{-}\rangle$ and $\langle\hat{b}\rangle=\sqrt{N_{\mathrm{b}}/p_{\mathrm{b}}}\langle\hat{\mathrm{k}}^{-}\rangle$, which associate the collective displacements with the mean spin of the ensembles.

\begin{figure*}
\centering{}\includegraphics[width=\textwidth]{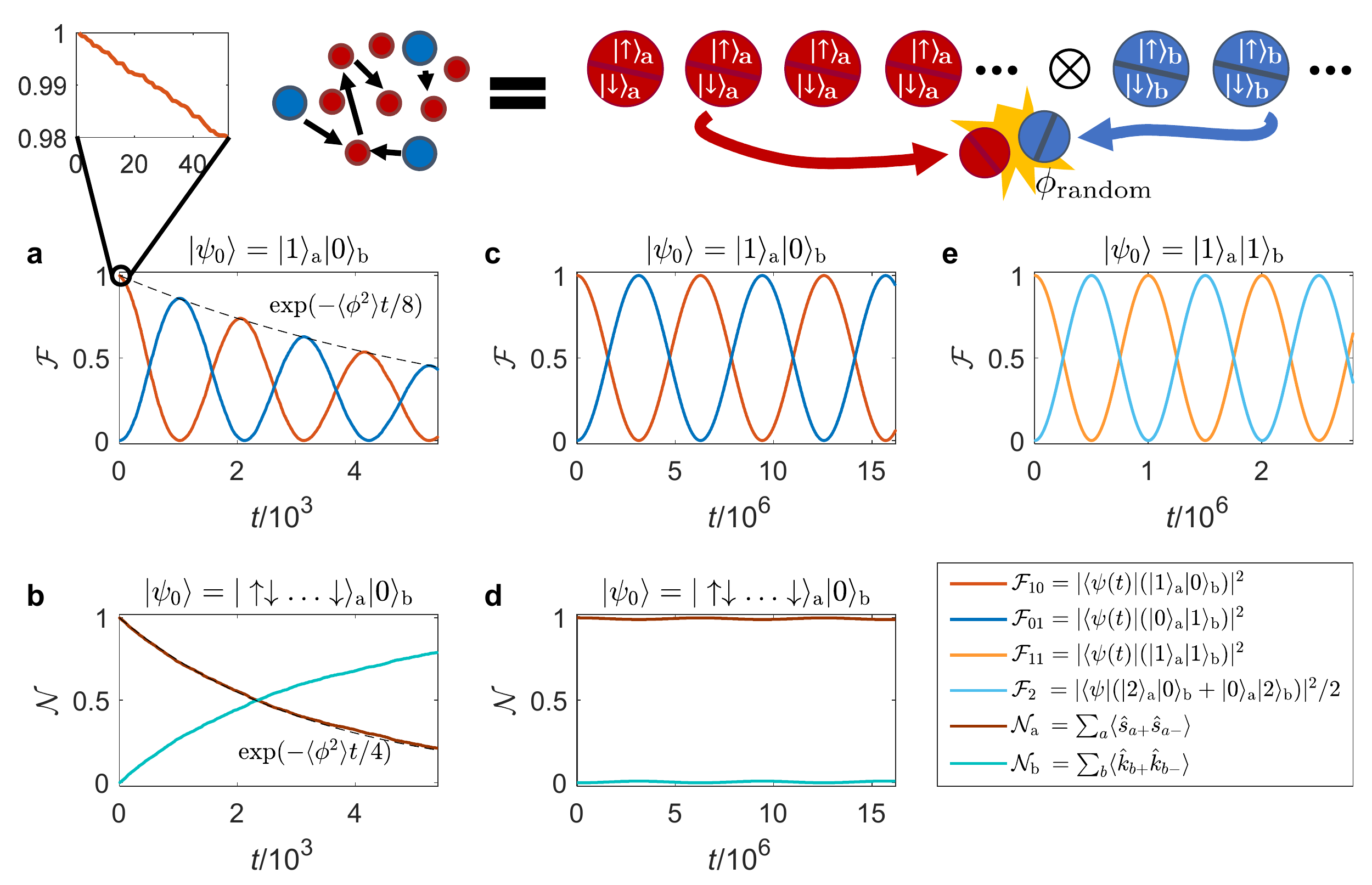}\caption{\textbf{Stochastic simulation of the collisional interface.}
We solve the unitary evolution of the quantum state of $N_{\mathrm{a}}=100$ electron spins and $N_{\mathrm{b}}=10^{4}$ noble-gas spins (in \textbf{e}, $N_{\mathrm{a}}=30$, $N_{\mathrm{b}}=300$), initialized in the state $|\psi_{0}\rangle$.
Each electron spin undergoes a spin-exchange collision of random strength ($\phi_\text{random}\ll1$) with a randomly chosen noble-gas spin every simulation time-step $\tau=1$; see SI for more details.
The collisions are either very weak $\langle\phi\rangle=10^{-5}$ (\textbf{c-e}) or more moderate $\langle\phi\rangle=2\cdot10^{-2}$ (\textbf{a,b}).
The electron spins can be initialized with all pointing downwards $|0\rangle_{\mathrm{a}}=|\downarrow\ldots\downarrow\rangle_{\mathrm{a}}$ (``vacuum''), or with one arbitrary spin pointing upwards $\hat{\mathrm{f}}_{a=1}^{+}|0\rangle_{\mathrm{a}}=|\uparrow\downarrow\ldots\downarrow\rangle_{\mathrm{a}}$ (``localized excitation''), or in the symmetric state with a single collective excitation $|1\rangle_{\mathrm{a}}=N_{\mathrm{a}}^{-1/2}\sum_{a}\hat{\mathrm{f}}_{a}^{+}|0\rangle_{\mathrm{a}}=N_{\mathrm{a}}^{-1/2}\sum_{i}|\downarrow\ldots\downarrow\uparrow_{i}\downarrow\ldots\downarrow\rangle_{\mathrm{a}}$ (``collective excitation'').
The nuclear spins are initialized in either $|0\rangle_{\mathrm{b}}$ or $|1\rangle_{\mathrm{b}}$.
\textbf{a}, An initial symmetric excitation $|\psi_{0}\rangle=|1\rangle_{\mathrm{a}}|0\rangle_{\mathrm{b}}$ is coherently exchanged between the two spin ensembles.
The inset highlights the stochasticity of the process.
The exchange is accompanied by dephasing due to incoherent transfer of the excitation to the large noble-gas ensemble ($N_{\mathrm{b}}\gg N_{\mathrm{a}}$) via the same process underlying SEOP.
\textbf{b}, The localized excitation is incoherently transferred to the noble gas.
\textbf{c}, Strikingly, when the collisions are weaker, the exchange fidelities $\mathcal{F}_{10}$ and $\mathcal{F}_{01}$ oscillate with higher contrast and nearly no decay, despite the stochasticity of the process.
\textbf{d}, Almost no oscillations are observed for the localized excitation.
\textbf{e}, When $|\psi_{0}\rangle=|1\rangle_{\mathrm{a}}|1\rangle_{\mathrm{b}}$, the two excitations periodically ``bunch'' in a superposition of either of the spin ensembles ($|2\rangle_{\mathrm{a}}|0\rangle_{\mathrm{b}}+|0\rangle_{\mathrm{a}}|2\rangle_{\mathrm{b}}$), manifesting the nonclassical Hong-Ou-Mandel phenomenon and validating the quantum beam-splitter property of the interface.
\label{fig:stochastic-simulation}}
\end{figure*}

\textbf{Coupling regimes.}
With the coupling rate $J$, detuning $\Delta$, and relaxation $\gamma$, Eqs.~(\ref{eq:coherent-dynamics}) has the canonical form of a coupled two-mode system \cite{Polzik2010ReviewRMP}.
While $J$ cannot be varied rapidly, $\Delta\left(B\right)$ can be controlled efficiently by varying the external magnetic field $B$ along the polarization axis.
$B$ alters $\Delta$ by predominantly altering the precession frequency of the alkali spins, owing to the $(100-1000)$-fold difference in the gyromagnetic ratios $g_{\mathrm{a}}$ and $g_{\mathrm{b}}$.
When the interaction is set off-resonance $|\Delta(B)|\gg J,\gamma$, the two collective spins effectively decouple.
This decoupling is often used in sensing applications to diminish the effect of the alkali on the noble-gas dynamics \cite{WalkerLarsen2016NGCNMRG,Safronova2018BudkerParticlesMagnetometry,Lee2018RomalisConstraints,Katz2021NobleSpectroscopy}.
In this regime, the alkali and noble-gas spins precess independently; the alkali spin experiences fast relaxation at a rate $\gamma$, while the noble-gas spin maintains its long coherence time.
We simulate these dynamics first for coherent spin states, as shown in Fig.~\ref{fig:coupling modes}a. 

Conversely, when the magnetic field is tuned to the so-called `compensation point' $\Delta(B)=0$ \cite{KornackRomalis2002OverlappingEnsemles}, the interaction becomes resonant, and the two spin ensembles hybridize.
The magnetic field thus acts as a controllable switch, rapidly coupling or decoupling the two spin ensembles.
Romalis and coworkers have demonstrated the alkali-noble-gas hybridization in the overdamped regime $\gamma\gtrsim J,|\Delta|$ \cite{KornackRomalis2002OverlappingEnsemles}.
In this regime, the noble-gas spins inherit a large fraction of the alkali spins' decoherence rate and thus thermalize before the transfer of excitations is complete, as shown in Fig.~\ref{fig:coupling modes}b.
The overdamped regime features a large enhancement in the sensitivity to various external fields and enables the operation as a comagnetometer sensor \cite{KornackRomalis2002OverlappingEnsemles,Kornack2005GyroComagRomalis}.

Here we focus on the recently demonstrated regime $J\gg\gamma$, which we identify as \textit{strong coupling} \cite{Shaham2021StrongCoupling}.
In this regime, the evolution is governed by the beam-splitter Hamiltonian $\hbar J(\hat{a}^{\dagger}\hat{b}+\hat{b}^{\dagger}\hat{a})$, which leads to the exchange of quantum states between the spin ensembles.
This is illustrated in Fig.~\ref{fig:coupling modes}c for coherent states, demonstrating a coherent transfer of spin excitations, as recently observed \cite{Shaham2021StrongCoupling}.
One can dynamically tune the exchange rate by varying the magnetic field strength.
In particular, maintaining the resonance conditions $\Delta(B)=0$ for a duration $t=\pi/(2J)$ and subsequently ramping $B$ up to $\Delta\gg J$ yields a deterministic state transfer between the two ensembles akin to a $\pi$ pulse.

The accumulated effect of the spin-exchange collisions is coherent only as long as their \emph{incoherent} contribution is negligible.
As described by the incoherent-transfer term in Eq.~(\ref{eq:Bloch-dynamics}), spin-exchange collisions introduce an additional relaxation rate $k_{\text{se}}n_{\mathrm{b}}/q$ into the alkali spin relaxation $\gamma$.
Consequently, since $n_{\mathrm{b}}\gg n_{\mathrm{a}}$, the strong coupling condition $J\gg\gamma$ requires that the precession angle $\langle\phi\rangle$ remains very small $\langle\phi\rangle\ll \sqrt{q p_{\mathrm{a}} p_{\mathrm{b}}} \sqrt{n_{\mathrm{a}}/n_{\mathrm{b}}}<1$.
Thus, strong coupling of the spin ensembles relies on the weakness of the individual spin-exchange collisions.

\textbf{Stochastic numerical simulations.}
To attest to the quantum nature of the interface, we develop a stochastic many-body simulation, which tracks the quantum state of many spins that randomly collide in the strong-coupling regime (see SI).
The simulation allows us to witness and visualize the coherent and collective outcome of the stochastic spin-exchange interaction as well as the associated relaxation.
For the sake of simplicity, we assume in the simulations $I=0$, \emph{i.e.} $\boldsymbol{\hat{\textbf{f}}}_{a}\equiv\boldsymbol{\hat{\textbf{s}}}_{a}$.
We initialize the system with either the symmetric excitation $|1\rangle_{\mathrm{a}}|0\rangle_{\mathrm{b}}\equiv N_{\mathrm{a}}^{-1/2}\sum_{a}\hat{\mathrm{f}}_{a}^{+}|0\rangle_{\mathrm{a}}|0\rangle_{\mathrm{b}}$ (Figs.~\ref{fig:stochastic-simulation}a,c) or a localized excitation $\hat{\mathrm{f}}_{a=1}^{+}|0\rangle_{\mathrm{a}}|0\rangle_{\mathrm{b}}$ (Figs.~\ref{fig:stochastic-simulation}b,d).
The exchange between the two ensembles emerges as a collective phenomenon: for the symmetric Fock state, we observe multiple, high-contrast oscillations of the populations of the collective states, whereas, for the localized excitation, the oscillations are negligible.
The transfer amplitude accumulates constructively only for the excitation that is symmetrically shared among all spins, maximizing the periodic exchange rate and fidelity.
This comparison emphasizes the collective nature of the coupling, which is taken for granted in the mean-field description.

As noted above, the dominance of the coherent exchange over the incoherent transfer and dephasing relies on the collisions being very weak.
To exemplify this, we compare between $\langle\phi\rangle=2\cdot10^{-2}$ (Figs.~\ref{fig:stochastic-simulation}a,b) and $\langle\phi\rangle=10^{-5}$ (Figs.~\ref{fig:stochastic-simulation}c,d), the latter corresponding to realistic $^{3}$He-potassium collisions.
For $\langle\phi\rangle=2\cdot10^{-2}$, we observe a dephasing of the alkali spin, with a rate $\gamma\propto k_{\text{se}}n_{\mathrm{b}}$, reducing the exchange fidelity compared to that with $\langle\phi\rangle=10^{-5}$.
Finally, we simulate the periodic bunching of two spin excitations, as shown in Fig.~\ref{fig:stochastic-simulation}e.
This evolution is analogous to the nonclassical Hong-Ou-Mandel phenomenon, demonstrating that the collisional interface supports the reversible, high fidelity, full exchange of nonclassical states between the two spin ensembles.

\begin{figure}[t]
\begin{centering}
\includegraphics[width=\columnwidth]{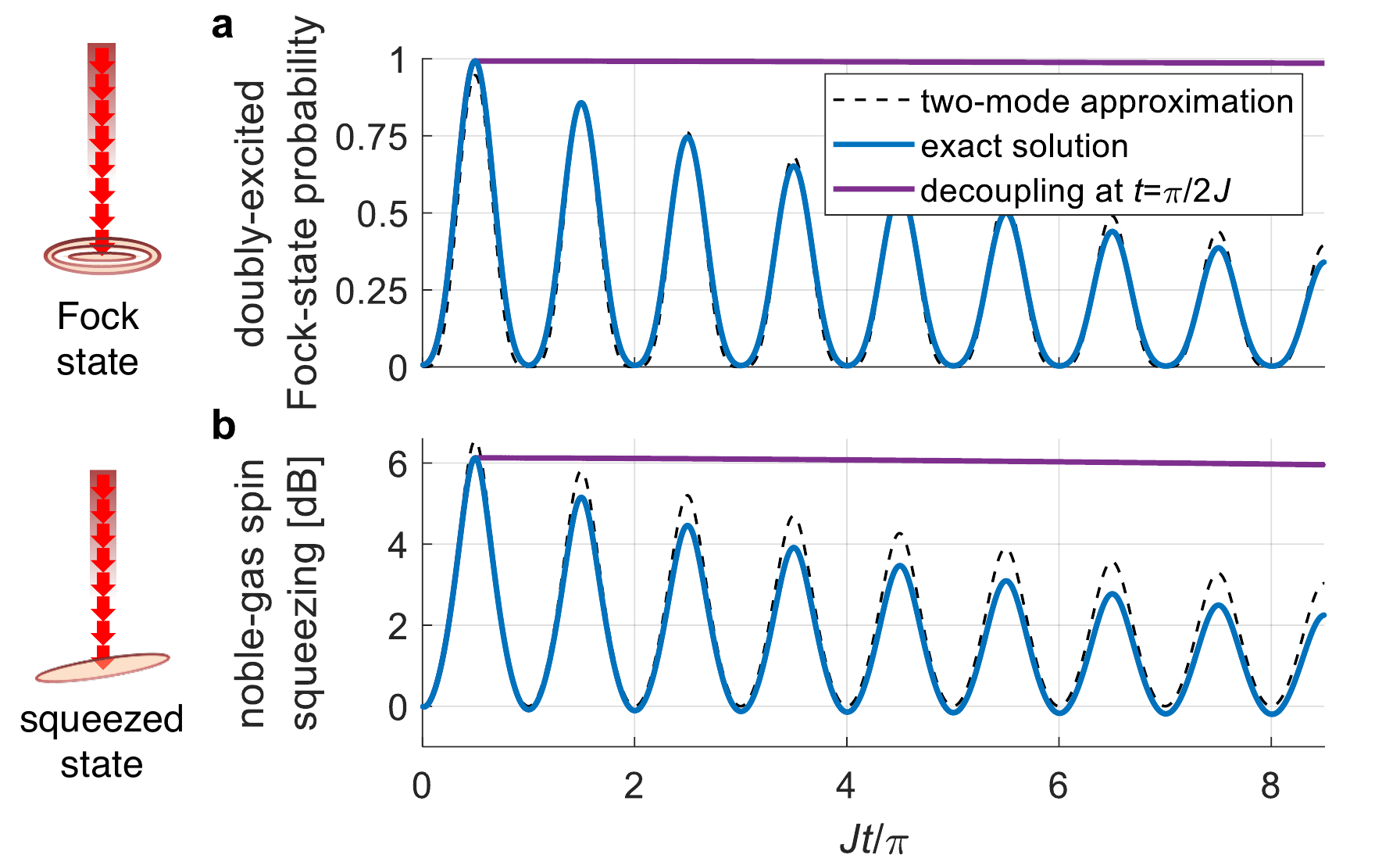}
\par\end{centering}
\centering{}\caption{\textbf{Exchange of nonclassical states between alkali and noble-gas
spin ensembles.}
The exchange is calculated for the proposed experimental parameters and assuming the initial excitation is spatially uniform.
We compare the two-mode approximation given by Eqs.~(\ref{eq:coherent-dynamics}) for perfectly-polarized ensembles (black) with an exact solution accounting for higher spatial modes and imperfect polarization (blue, see Methods).
At $t=0$, the magnetic field is tuned to resonance $\Delta\left(B\right)=0$.
If it is detuned at $t=\pi/(2J)$, the state transfer is maximal (solid purple, using $B=180\,\mathrm{mG}$). The imperfect polarization is accounted for by initializing the system with an incoherent state and incorporating the associated excess quantum noise (see Methods).
\textbf{a}, Probability of populating the uniform (longest-lived) diffusion mode of the noble gas with exactly two excitations (Fock state) when the system is initiated with two excitations in alkali's uniform mode.
\textbf{b}, Exchange of squeezing between the total alkali spin (proportional to the electron spin $\hat{S}_{x}=\sum_{a}\hat{\text{s}}_{a,x}$) and the total noble-gas spin ($\hat{K}_{y}=\sum_{b}\hat{\text{k}}_{b,y}$).
The alkali is initialized with $7\,\mathrm{dB}$ squeezing.
\label{fig:excitation exchange fidelity}}
\end{figure}

To elucidate the physical mechanism that renders a deterministic exchange out of random collisions, we analytically describe the quintessential case of an exchange of a single spin excitation.
The noble-gas spins are initialized in the state $|0\rangle_{\mathrm{b}}$ with all spins pointing down, and the alkali spins are initialized in the nonclassical Fock state $|1\rangle_{\mathrm{a}}\equiv N_{\mathrm{a}}^{-1/2}\sum_{a}\hat{\mathrm{f}}_{a}^{+}|0\rangle_{\mathrm{a}}$, \emph{i.e.,} a symmetric superposition with one of the spins pointing up.
After a short time $t$, the system wavefunction $|1\rangle_{\mathrm{a}}|0\rangle_{\mathrm{b}}$ evolves into 
\begin{equation}
|1\rangle_{\mathrm{a}}|0\rangle_{\mathrm{b}}-iJt|0\rangle_{\mathrm{a}}|1\rangle_{\mathrm{b}}-i\epsilon|\delta\psi\rangle+\mathcal{O}(\langle\phi^{2}\rangle).
\label{eq:spin-detail}
\end{equation}
This evolution, with $Jt,\epsilon\ll1$, is the onset of transfer of the single spin excitation from the alkali to the noble gas via both deterministic and stochastic contributions; The Fock state $|1\rangle_{\mathrm{b}}\equiv N_{\mathrm{b}}^{-1/2}\sum_{b}\hat{\mathrm{k}}_{b}^{+}|0\rangle_{\mathrm{b}}$ manifests the deterministic transfer, while the \textit{stochastic wavefunction} $|\delta\psi\rangle$ represents an incoherent mixture of excited spins (see SI).
To quantify the corresponding transition amplitudes $Jt$ and $\epsilon$,  we assume that any alkali is equally likely to collide with any noble-gas atom at a mean time between collisions $\tau=1/(n_{\mathrm{b}}\sigma v)$.
We then find that $Jt=\tfrac{1}{2}(\langle\phi\rangle t/\tau)\sqrt{N_{\mathrm{a}}/N_{\mathrm{b}}}$ and that, after many collisions, the stochastic variable $\epsilon$ follows the central limit theorem $\epsilon\rightarrow\sqrt{\langle\phi^{2}\rangle t/(2\tau)}$.
These results show that, at longer times, the transition amplitude of the deterministic term in Eq.~(\ref{eq:spin-detail}) dominates the transition amplitude of the stochastic term, since $Jt\gg\epsilon$.
It is the accumulated mean effect of many weak collisions that leads to deterministic transfer scaling linearly with $t/\tau$ and to fluctuations that only add up incoherently as $\sqrt{t/\tau}$.

\textbf{Outline for a non-classical demonstration.}
We now identify relevant experimental parameters for realizing strong coupling in a mixture of potassium-39 and helium-3.
To maximize the collective coupling rate $J$, we consider reasonably high densities of $n_{\mathrm{a}}=3\times10^{14}\,\text{\ensuremath{\unit{cm^{-3}}}}$ (vapor pressure at $215^{\circ}\text{C}$), $n_{\mathrm{b}}=2\times10^{20}\,\text{\ensuremath{\unit{cm^{-3}}}}$ (7.5~atm), and 30$\,$Torr of $\mathrm{N}_{2}$ for quenching \cite{KornackRomalis2002OverlappingEnsemles}.
Using standard optical pumping, the alkali spin polarization can be initialized in a spin-temperature distribution with $p_{\mathrm{a}}\geq0.95$, for which the slowing down factor is $q=4.1$ \cite{Polzik2010ReviewRMP,Sheng2013RomalisSubFemtoTesla}.
The noble-gas spin can be initialized via SEOP to a moderate yet sufficient polarization of $p_{\mathrm{b}}\gtrsim0.75$ \cite{ChenBabcockWalker2014HigherLimitsSEOPHe3}.

A coupling rate of $J=1000\,\text{s}^{-1}$ is reached at this temperature, as $v\sigma\langle\phi\rangle=2\times10^{-14}\,\text{\ensuremath{\unit{cm^{3}}}}\text{s}^{-1}$ (corresponding to $\langle\phi\rangle^{2}\approx\langle\phi^{2}\rangle\approx2\times10^{-10}\,\text{rad}^{2}$).
The resonance condition $\Delta=(g_{\text{a}}-g_{\text{b}})B+\Delta_{\text{c}}=0$ is obtained for a magnetic field $B=94\,\text{mG}$, predominantly compensating for the large collisional shift $v\sigma\langle\phi\rangle p_{\mathrm{b}}n_{\mathrm{b}}/2q$ experienced by the potassium, and yielding the Larmor frequency $g_{\text{b}}B=g_{\text{a}}B+\Delta_{\text{c}}=300\,(2\pi)\,\mathrm{Hz}$.
The high alkali density and polarization and the relatively small Larmor frequency puts the potassium spins in the 'spin-exchange relaxation-free' (SERF) regime \cite{Savukov2005RomalisSERF,Kominis2003RomalisSERFmagnetometer,Katz2013nonlinearSERF}, rendering their relaxation via spin-exchange collisions negligible.
The relaxation rate is governed by spin-rotation interaction with $^{3}\text{He}$ and $\text{N}_{2}$ and by spin-destruction collisions with other potassium atoms, giving $\gamma=17.5\,\text{\ensuremath{\text{s}^{-1}}}$
\cite{Happer2010book}.
We thus reach the strong-coupling regime with potentially $J>55\gamma$.

The spin state of $^{3}\text{He}$ in this system can endure for $100$ hours, providing that magnetic-field gradients, magnetic impurities in the cell, and alkali-induced dephasing are minimized \cite{Walker2017He3review,Walker1997SEOPReview,Limes2018HeXecomag}.
The alkali spins can be initialized in a nonclassical state via entanglement-generation schemes \cite{Julsgaard2001PolzikEntanglement,Vasilakis2015PolzikBackactionEvation} or by mapping nonclassical light onto the spin orientation moment \cite{katz2018storage1sec}.
Calculations of the spin dynamics with these parameters in a $2"$-diameter spherical cell are presented in Fig.~\ref{fig:excitation exchange fidelity}.
These calculations account for the spatial dynamics and for the initial imperfect polarization by employing a non-pure density matrix with multiple diffusion modes \cite{Shaham2020Diffusion}, initialized with incoherent spin excitations corresponding to $p_{\mathrm{a}}=0.95$ and $p_{\mathrm{b}}=0.75$.
We include the increased quantum noise due to imperfect polarization and calculate expectation values by tracing over the contribution of all diffusion modes (see Methods).
These calculations demonstrate the exchange of nonclassical Fock state and squeezed state and their mapping onto the long-living noble-gas spins in a realistic regime of imperfect spin-polarization.
\\

\textbf{Discussion.}
We present analytic and numeric  quantum-mechanical models for the hybrid system of alkali-metal and noble-gas spins.
The models reveal a collective mechanism that couples the macroscopic quantum states of the two spin ensembles.
We highlight feasible experimental parameters for reaching the strong-coupling regime, which enables a faithful quantum-state transfer between the alkali and noble-gas ensembles.

It is intriguing that weak collisions, despite their random nature, allow for an efficient, reversible, and controllable exchange of excitations.
It is particularly counter-intuitive that this exchange preserves the unique quantum statistics of non-classical states.
Equations (\ref{eq:coherent-dynamics}) manifest a genuine quantum interface, as they describe the exchange between the operators $\hat{a}$ and $\hat{b}$, which in turn encapsulate the full quantum statistics of the collective spins states.
The effect of the randomness of collisions on the quantum statistics is then incorporated in the noise operator $\hat{F}_{\mathrm{a}}$.

In stochastic quantum systems, the variance of quantum noise satisfies $\langle\hat{F}_{\mathrm{a}}\hat{F}_{\mathrm{a}}^{\dagger}\rangle\geq2\gamma$ for any relaxation rate $\gamma$, where equality is obtained for the case of vacuum noise \cite{GardinerZoller2004quantumNoise}.
For perfect spin polarization, we find that the noise due to spin-exchange collisions is a vacuum noise, that is, the minimal possible for an open quantum system (see Methods).
This result is apparent, for example, in the exchange of a single spin excitation between perfectly-polarized ensembles, where we obtain $\langle\hat{F}_{\mathrm{a}}\hat{F}_{\mathrm{a}}^{\dagger}\rangle=\epsilon^{2}/t=2\gamma$.

While it is experimentally possible to approach unity polarization of the alkali atoms $p_{\mathrm{a}}\rightarrow 1$ \cite{KarstenPolzik2020SinglePhoton,Chann2003BabcockWalker75He3}, the highest $^{3}$He polarization demonstrated to date is $p_{\mathrm{b}}=0.85$ \cite{ChenBabcockWalker2014HigherLimitsSEOPHe3}.
It is therefore important to discuss the consequences of imperfect spin polarization $p_{\mathrm{a}},p_{\mathrm{b}}<1$.
The first and more trivial consequence is a moderate reduction of $J\propto\sqrt{p_{\mathrm{a}}p_{\mathrm{b}}}$, since now only a fraction $p_{\mathrm{a}}p_{\mathrm{b}}$ of the atomic collisions contribute to the collective exchange process.
The second consequence is an added noise due to initial incoherent population of transverse spin excitations.
These incoherent excitations are distributed over a macroscopic number of spatial modes.
Therefore, despite having a macroscopic number of depolarized atoms, the number of excitations per mode can be small.
Notably, for a finite polarization degree $0<p\le 1$, the mean number of incoherent excitations in the uniform spatial mode is $\bar{n}_\mathrm{I}=(1-p)/2p$.
For $p$ close to unity, $\bar{n}_\mathrm{I}\ll1$ is small.
Therefore these excitations might contribute a weak classical background, in principle relevant at all times $t>0$, which reduces the fidelity of non-classical states.
For example, there is a non-zero probability $\bar{n}_\mathrm{I}\ll1$ of finding a spin excitation without any stimulation.
Finally, the incoherent excitations residing in all other modes form a `thermal reservoir' of spins that is manifested as an excess quantum noise.
In particular, the collisional coupling of the collective spin to this reservoir increases the variance of the quantum noise operators acting on the spin.
This noise is accumulated during the dynamics and is thus less relevant at short time scales.
Importantly, the excess noise is moderate and, for highly-polarized ensembles, it is on the same order as the vacuum noise.

The quantum interface we study allows for a controllable state-exchange between two spin-gas species, and it is of particular importance when it comes to noble-gas spins, which are extremely long-lived but optically inaccessible.
The interface can reach the strong coupling regime and allows for non-adiabatic exchange, thus significantly increasing operations bandwidth.
In conjunction with recent experiments demonstrating the coherent, efficient, and bi-directional properties of the collective coupling in the classical regime \cite{Shaham2021StrongCoupling,Katz2021NobleSpectroscopy}, our study thus opens a path to couple light to the transparent spins in the quantum regime.
The scenario resembles quantum-logic operations with nuclear ensembles in solids, where a long-lived nuclear spin is accessible via the hyperfine interaction with an electron spin, which is optically manipulated and interrogated \cite{Gangloff2019QDNuclearSpinInterface}.
The spin-exchange interface therefore paves the way towards wider applications of noble-gas spins in quantum optics, including long-lived quantum memories and long-distance entanglement at ambient conditions \cite{AlkaliNobleEntanglementKatz2020PRL,noblegasStoragePRA2020arxiv,noblegasStoragePRL2020arxiv}, as well as to fundamental research of the limits of quantum theory for entangled macroscopic objects.

\newpage

\part*{\centerline{Methods}}

\section*{Detailed model of the Many-Body problem}

\textbf{Spin-exchange interaction.}
Consider a gaseous mixture of $N_{\mathrm{a}}$ alkali-metal atoms and $N_{\mathrm{b}}$ noble-gas atoms enclosed in a spherical cell.
Each alkali atom, labeled by $a$, has a valence electron with a spin $S=1/2$ operator $\boldsymbol{\hat{\textbf{s}}}_{a}$, in addition to its nuclear spin $I>0$ with spin operator $\boldsymbol{\hat{\textbf{i}}}_{a}$.
Each noble-gas atom, labeled by $b$, has a nucleus with a $K=1/2$ spin, represented by $\boldsymbol{\hat{\textbf{k}}}_{b}$.
During collisions, the noble-gas spin interacts only with the valence electron.
The Hamiltonian of the two spin ensembles is given by $\mathcal{H}\left(t\right)=\mathcal{H}_{0}+\mathcal{V}\left(t\right)$, where 
\begin{equation}
\mathcal{H}_{0}=\hbar a_{\text{hpf}}\sum_{a}\boldsymbol{\hat{\textbf{i}}}_{a}\boldsymbol{\hat{\textbf{s}}}_{a}+\hbar{\normalcolor \tilde{\omega}_{a}q}\sum_{a}\hat{\text{s}}_{a,z}+\hbar\tilde{\omega}_{\textnormal{b}}\sum_{b}\hat{\text{k}}_{b,z}\label{eq:H0_hamiltonian}
\end{equation}
is the non-interacting Hamiltonian of the two spin ensembles.
$a_{\text{hpf}}$ denotes the hyperfine coupling constant in the ground state of the alkali atom \cite{Happer2010book}, and $\tilde{\omega}_{\mathrm{a}}$ and $\tilde{\omega}_{\mathrm{b}}$ are the Larmor frequencies of the alkali and noble-gas spins induced by an external magnetic field $\mathbf{B}=B\hat{z}$.
The microscopic many-body interaction Hamiltonian, governed by the Fermi-contact interaction \cite{Walter1998HapperWalkerPhiTrajectory,Romalis2014CommentGradientsSphere}, is given by 
\begin{equation}
\mathcal{V}(t)=\sum_{a=1}^{N_{\mathrm{a}}}\sum_{b=1}^{N_{\mathrm{b}}}\hbar\alpha_{ab}(t)\boldsymbol{\hat{\textbf{s}}}_{a}\cdot\boldsymbol{\hat{\textbf{k}}}_{b}.
\end{equation}
This form conserves the total spin of the colliding pairs.
The instantaneous interaction strength $\alpha_{ab}\left(t\right)$ between atoms $a$ and $b$ is determined by the specific microscopic trajectory of each atom. 

The spatial degrees of freedom of the thermal atoms are classical.
Their coordinates $\mathbf{r}_{a}\left(t\right)$ and $\mathbf{r}_{b}\left(t\right)$ follow ballistic trajectories, which are independent of the spin state and governed by the classical Langevin equation.
The collisions in the gas can be considered as sudden and binary; the mean collision duration $\tau_\text{c}$ is only a few picoseconds \cite{Happer2010book}, whereas the mean time between collisions for an alkali atom $\tau$ is a few nanoseconds at ambient pressure.
Since collisions are isolated in time $(\tau_\text{c}\ll\tau)$, the interaction strength can be approximated by a train of instantaneous events $\alpha_{ab}\left(t\right)=\sum_{i}\hbar\phi_{ab}^{\left(i\right)}\delta(t-t_{ab}^{\left(i\right)})$.
Here $\phi_{ab}^{\left(i\right)}$ denotes the phase $\phi$ that spins $a$ and $b$ accumulate during the $i^{\text{th}}$ collision, and $t_{ab}^{\left(i\right)}$ denotes the time of collision, as determined from the particles trajectories (see SI).

We consider short times $\tau'$, typically a few tens of picoseconds, such that $\tau\gg\tau'\gg\tau_\text{c}$, for which each atom experiences at most a single collision.
In other words, we assume that if a collision occurred between an alkali spin $a$ and a noble-gas spin $b$, then neither $a$ nor $b$ collided with other atoms during $\tau'$.
Consequently, $\mathcal{V}\left(t\right)$ has no more than one appearance of each spin operator and thus commutes with itself.
Under these conditions, the time-evolution operator is simplified to 
\begin{equation}
U_\text{I}\left(t+\tau',t\right)=\exp\Bigl(-i\sum_{ab}\sum_{i}^{\tau'}\phi_{ab}^{\left(i\right)}\boldsymbol{\hat{\textbf{s}}}_{a}\cdot\boldsymbol{\hat{\textbf{k}}}_{b}\Bigr).\label{eq:Dyson-series-collapse}
\end{equation}
Here $\sum_{i}^{\tau'}$ denotes the sum over all collision instances that occur during the short time interval in which $t_{ab}^{\left(i\right)}\in\left[t,t+\tau'\right]$.
For weak collisions, the mutual precession is small $\phi_{ab}^{\left(i\right)}\ll1$, and the exponential term in Eq.~(\ref{eq:Dyson-series-collapse}) can be expanded to leading orders in $\phi$ as a Dyson series
\begin{equation}
U_\text{I}\approx U_\text{I}^{\left(0\right)}+U_\text{I}^{\left(1\right)}+U_\text{I}^{\left(2\right)}+\ldots\label{eq:Dyson-Series}
\end{equation}
Here the lowest-order terms are $U_\text{I}^{\left(0\right)}\left(t+\tau',t\right)=\mathds{1}$ (the identity),
\[
\begin{aligned}U_\text{I}^{\left(1\right)}\left(t+\tau',t\right) & =-i\sum_{ab}\sum_{i}^{\tau'}\phi_{ab}^{\left(i\right)}\boldsymbol{\hat{\textbf{s}}}_{a}\cdot\boldsymbol{\hat{\textbf{k}}}_{b},\mathrm{\,\,\,\,\,\,\,and}\\
U_\text{I}^{\left(2\right)}\left(t+\tau',t\right) & =-\frac{1}{2}\sum_{abcd}\sum_{ij}^{\tau'}\phi_{ab}^{\left(i\right)}\phi_{cd}^{\left(j\right)}(\boldsymbol{\hat{\textbf{s}}}_{a}\cdot\boldsymbol{\hat{\textbf{k}}}_{b})(\boldsymbol{\hat{\textbf{s}}}_{c}\cdot\boldsymbol{\hat{\textbf{k}}}_{d}).
\end{aligned}
\]
This simplified form provides for the evolution of any quantum mechanical operator $\hat{A}$ after time $\tau'$, $\Delta\hat{A}=U^{\dagger}\left(t+\tau',t\right)\hat{A}\left(t\right)U\left(t+\tau',t\right)-\hat{A}\left(t\right)$, where $U=e^{-i\mathcal{H}_0\tau'}U_\text{I}$ is in the Heisenberg picture.
Note that $e^{-i\mathcal{H}_0\tau'}$ and $U_\text{I}$ commute, as explained below.
Up to second order in $\phi$, the dynamics of $\hat{A}$ is given by 
\begin{align}
\frac{1}{\tau'}\Delta\hat{A} & =-\frac{i}{\hbar}[\hat{A},\mathcal{H}_{0}]-\frac{i}{\hbar}[\hat{A},\mathcal{V}]+\mathcal{L}\bigl(\hat{A}\bigr).\label{eq:discrete-step-evolution}
\end{align}
The first term is the standard Hamiltonian evolution governed by $\mathcal{H}_{0}$ and independent of $\phi$.
The second term describes a unitary evolution during a collision with an effective Hamiltonian
\begin{equation}
\mathcal{V}=\sum_{ab}\sum_{i}^{\tau'}\frac{\hbar}{\tau'}\phi_{ab}^{\left(i\right)}\boldsymbol{\hat{\textbf{s}}}_{a}\cdot\boldsymbol{\hat{\textbf{k}}}_{b},\label{eq:coarse-grained-Hamiltonian}
\end{equation}
which is first-order $\phi$.
The third term $\mathcal{L}\left(A\right)$ is proportional to $\phi^{2}$ and has the structure of a standard Lindblad term
\begin{equation}
\mathcal{L}\left(A\right)=-\cfrac{1}{2}\sum_{ab}\sum_{i}^{\tau'}(\phi_{ab}^{\left(i\right)})^{2}\Bigl[\boldsymbol{\hat{\textbf{s}}}_{a}\cdot\boldsymbol{\hat{\textbf{k}}}_{b},\bigl[\boldsymbol{\hat{\textbf{s}}}_{a}\cdot\boldsymbol{\hat{\textbf{k}}}_{b},\hat{A}\bigr]\Bigr].\label{eq:La}
\end{equation}
We note however, that this operator is not associated with a decay but is rather a second-order correction to the unitary evolution.

To describe the short-time evolution of the spin ensembles, we derive the time-evolution operator $U_\text{I}$ in the interaction picture.
This operator satisfies $i\hbar\partial_{t}U_\text{I}=\mathcal{V}_\text{I}U_\text{I}$, where $\mathcal{V}_\text{I}\left(t\right)=e^{\frac{i}{\hbar}\mathcal{H}_{0}t}\mathcal{V}\left(t\right)e^{-\frac{i}{\hbar}\mathcal{H}_{0}t}$ is the Hamiltonian in the interaction picture.
The collisions are sudden $\mathcal{H}_{0}\tau_\text{c}\lll1$ (except at strong magnetic fields greater then tens of Tesla), rendering the evolution by $\mathcal{H}_{0}$ negligible during the short time of collision (typically $\tau_\text{c}\approx1\,\text{psec}$ and $B<1\unit{G}$, such that $\tilde{\omega}_{a}\tau_\text{c}\lesssim10^{-7}$).
As a result, we can assume that $\mathcal{H}_{0}$ and $\mathcal{V}$ commute, hence $\mathcal{V}_\text{I}\left(t\right)=\mathcal{V}\left(t\right)$.

The evolution of the single-spin operators $\boldsymbol{\hat{\textbf{s}}}_{a}$ and $\boldsymbol{\hat{\textbf{k}}}_{b}$ in the time interval $\tau'$ are then derived from Eq.~(\ref{eq:discrete-step-evolution}), yielding
\begin{equation}
\begin{array}{cc}
\Delta\boldsymbol{\hat{\textbf{s}}}_{a} & =\sum_{n}\sum_{i}^{\tau'}\phi_{an}^{\left(i\right)}[\boldsymbol{\hat{\textbf{k}}}_{n}\times\boldsymbol{\hat{\textbf{s}}}_{a}+\phi_{an}^{\left(i\right)}(\boldsymbol{\hat{\textbf{k}}}_{n}-\boldsymbol{\hat{\textbf{s}}}_{a})/4],\\
\Delta\boldsymbol{\hat{\textbf{k}}}_{b} & =\sum_{m}\sum_{i}^{\tau'}\phi_{mb}^{\left(i\right)}[\boldsymbol{\hat{\textbf{s}}}_{m}\times\boldsymbol{\hat{\textbf{k}}}_{b}+\phi_{mb}^{\left(i\right)}(\boldsymbol{\hat{\textbf{s}}}_{m}-\boldsymbol{\hat{\textbf{k}}}_{b})/4].
\end{array}\label{eq:spin-operators-discrete-step-evolution1}
\end{equation}
This form conserves the total spin of each colliding pair $a-b$, since $\Delta\bigl(\boldsymbol{\hat{\textbf{s}}}_{a}+\boldsymbol{\hat{\textbf{k}}}_{b}\bigr)=0$.
Equation (\ref{eq:spin-operators-discrete-step-evolution1}) describes the mutual precession of pairs of spins, as illustrated in Fig.~\ref{fig:illustration}a.
This evolution is unitary to second order in the precession angle $\phi$, while higher-order contributions are neglected in the truncation of Eq.~(\ref{eq:Dyson-Series}).

Between collisions, the nuclear spin of the \emph{alkali} atoms is altered by the strong hyperfine interaction with the electron.
Consequently, the slow dynamics of the alkali atoms should be described in terms of the operator sum $\boldsymbol{\hat{\textbf{f}}}_{a} =  \boldsymbol{\hat{\textbf{s}}}_{a} + \boldsymbol{\hat{\textbf{i}}}_{a}$.
Here we focus on alkali ensembles in a spin-temperature population-distribution, for which $\boldsymbol{\hat{\textbf{f}}}_{a}=q\boldsymbol{\hat{\textbf{s}}}_{a}$, with the slowing-down factor $q = q(I,p_\mathrm{a})$ given by \cite{Vasilakis2011dissertation}
\begin{equation}
q(I,p_\mathrm{a}) = \frac{2I+1}{p_\mathrm{a}} \frac{(1+p_\mathrm{a})^{2I+1}+(1-p_\mathrm{a})^{2I+1}}{(1+p_\mathrm{a})^{2I+1}-(1-p_\mathrm{a})^{2I+1}} - \frac{1}{p_\mathrm{a}^2} + 1. \label{eq:slowdown_factor}
\end{equation} 

The slow evolution of the spins depends on the cumulative effect of multiple collisions among different atoms.
At the macroscopic limit, it is formidable to keep track of the kinematic details of all atoms, given a large set of collision times $t_{ab}^{\left(i\right)}$ and strengths $\phi_{ab}^{\left(i\right)}$.
Instead, we represent the exact values of $t_{ab}^{\left(i\right)}$ and $\phi_{ab}^{\left(i\right)}$ by their equivalent random variables $\sum_{i}^{\tau'}\phi_{ab}^{\left(i\right)}\rightarrow\varkappa_{ab}(t,\tau')\phi_{a}(t)$ and $\sum_{i}^{\tau'}(\phi_{ab}^{\left(i\right)})^{2}\rightarrow\varkappa_{ab}(t,\tau')\phi_{a}^{2}(t)$.
Here $\varkappa_{ab}\left(t,\tau'\right)$ is a Bernoulli process indicating whether a collision between particles $a,b$ has occurred during the short time interval $[t,t+\tau']$, with $\tau'\gg\hbar/a_{\text{hpf}}$.
As the phase $\phi_{a}$ depends on the kinematic parameters of the collision, such as the impact parameter and the two-body reduced-velocity \cite{Walker1997SEOPReview}, we treat it as a random variable, with a mean $\langle\phi\rangle$ and variance $\mathrm{var}(\phi)$.
The operation $\langle\cdot\rangle$ denotes an average over the microscopic kinematic parameters.
The stochastic nature of $\phi_{a}$ manifests the randomness in the interaction strength, while the stochastic nature of $\varkappa_{ab}$ manifests the randomness in pairing the colliding atoms.
We derive the statistical properties of $\varkappa_{ab}$ as a function of the microscopic kinematic variables in the SI, yielding $\bigl\langle\varkappa_{ab}(t,\tau')\varkappa_{cd}(t',\tau')\bigr\rangle=\delta_{ac}\delta_{bd}\tau'\delta(t-t')\bigl\langle\varkappa_{ab}(t,\tau')\bigr\rangle$ and $\bigl\langle\varkappa_{ab}(t,\tau')\bigr\rangle=v\sigma\tau'w(\mathbf{r}_{a}-\mathbf{r}_{b})$.
Here the window function $w\left(\mathbf{r}\right)=\Theta(l-|\mathbf{r}|)/V_l$ represents a control volume $V_l=4\pi l^{3}/3$, where $\Theta$ is the Heaviside function, and $l$ is the coarse-graining scale (larger than the atoms mean free path, see SI).

We are now set to perform spatial coarse-graining.
First, we replace the discrete atomic operators with the continuous operators $\boldsymbol{\hat{\textbf{f}}}\left(\mathbf{r},t\right)\equiv\sum_{a}\boldsymbol{\hat{\textbf{f}}}_{a}\left(t\right)\delta[\mathbf{r}-\mathbf{r}_{a}\left(t\right)]$ and $\boldsymbol{\hat{\textbf{k}}}\left(\mathbf{r},t\right)\equiv\sum_{b}\boldsymbol{\hat{\textbf{k}}}_{b}\left(t\right)\delta[\mathbf{r}-\mathbf{r}_{b}\left(t\right)]$ (cf.~\cite{Polzik2010ReviewRMP,Fleischhauer2000LukinEITPolaritons}).
We then perform the spatial convolutions $\boldsymbol{\hat{\textbf{f}}}(\mathbf{r},t)\rightarrow\boldsymbol{\hat{\textbf{f}}}(\mathbf{r},t)\ast w\left(\mathbf{r}\right)$ and $\boldsymbol{\hat{\textbf{k}}}(\mathbf{r},t)\rightarrow\boldsymbol{\hat{\textbf{k}}}(\mathbf{r},t)\ast w\left(\mathbf{r}\right)$.
The central-limit theorem is valid for this coarse-graining operation as long as $V_l$ contains a large number of particles $V_ln_{\mathrm{a}},\,V_ln_{\mathrm{b}}\gg1$.
Consequently, $\boldsymbol{\hat{\textbf{f}}}\left(\mathbf{r},t\right)$ and $\boldsymbol{\hat{\textbf{k}}}\left(\mathbf{r},t\right)$ become local symmetric spin operators, and the spatial coordinate $\mathbf{r}$ supersedes the specific particle labels.

Next, we consider the collisional part of the evolution of $\boldsymbol{\hat{\textbf{f}}}(\mathbf{r},t)$ and $\boldsymbol{\hat{\textbf{k}}}(\mathbf{r},t)$ at time intervals much longer than $\tau',$
\begin{equation}
\begin{aligned}\partial_{t}\boldsymbol{\hat{\textbf{f}}} & =\zeta\boldsymbol{\hat{\textbf{k}}}\times\boldsymbol{\hat{\textbf{f}}}-k_{\text{se}}n_{\mathrm{b}}\boldsymbol{\hat{\textbf{f}}}+qk_{\text{se}}n_{\mathrm{a}}\boldsymbol{\hat{\textbf{k}}}+\boldsymbol{\hat{F}}_{\mathrm{ex}},\\
\partial_{t}\boldsymbol{\hat{\textbf{k}}} & =\zeta\boldsymbol{\hat{\textbf{f}}}\times\boldsymbol{\hat{\textbf{k}}}-k_{\text{se}}n_{\mathrm{a}}\boldsymbol{\hat{\textbf{k}}}+\tfrac{1}{q}k_{\text{se}}n_{\mathrm{b}}\boldsymbol{\hat{\textbf{f}}}-\boldsymbol{\hat{F}}_{\mathrm{ex}}.
\end{aligned}
\label{eq:coarse-graind_equation_dJ_dt}
\end{equation}
The first term in Eqs.~(\ref{eq:coarse-graind_equation_dJ_dt}) represents the average mutual precession of the two symmetric spin operators within the coarse-graining volume, with the local interaction strength given by $\zeta\equiv\left\langle \sigma v\phi\right\rangle /q$.
As it describes coherent dynamics, it can be associated with an effective spin-exchange Hamiltonian 
\begin{equation}
\mathcal{V}_{\mathrm{ex}}=\hbar\zeta\int d^{3}\mathbf{r}_{\mathrm{a}}\int d^{3}\mathbf{r}_{\mathrm{b}}\delta(\mathbf{r}_{\mathrm{a}}-\mathbf{r}_{\mathrm{b}})\boldsymbol{\hat{\textbf{f}}}(\mathbf{r}_{\mathrm{a}},t)\cdot\boldsymbol{\hat{\textbf{k}}}(\mathbf{r}_{\mathrm{b}},t).\label{eq:Coarse-grained-interaction-Hamiltonian}
\end{equation}
The second and third terms in Eqs.~(\ref{eq:coarse-graind_equation_dJ_dt}) represent incoherent transfer of spin polarization from one specie to the other, while conserving the total spin.
Recall that we assume a spin-1/2 noble gas.
Here $n_{\mathrm{a}}=\sum_{a}w(\mathbf{r}-\mathbf{r}_{a})$ and $n_{\mathrm{b}}=\sum_{b}w(\mathbf{r}-\mathbf{r}_{b})$ denote the local densities of the two spin ensembles, and $k_{\text{se}}\equiv\frac{1}{4}\bigl\langle v\sigma\phi^{2}\bigr\rangle$ is known as the binary spin-exchange rate coefficient \cite{Happer2010book}.
In particular, the term $k_{\text{se}}n_{\mathrm{b}}\boldsymbol{\hat{\textbf{f}}}/q$ is responsible for hyperpolarization of the noble gas by optically pumped alkali-metal spins, underlying the SEOP technique \cite{Happer2010book,Walker1997SEOPReview}.
The incoherent SEOP term here replaces the coherent $\mathcal{L}\left(A\right)$ defined in Eq.~(\ref{eq:La}); it has the same functional form but is now essentially incoherent due to the coarse-graining of the microscopic kinematics.
Notably, incoherent effects are second order in $\phi$, and since $\langle\phi^{2}\rangle/\langle\phi\rangle\lll1$ (typically $10^{-5}$), $\zeta$ is substantially larger than $k_{\mathrm{se}}$.
The fluctuation vector-operator $\boldsymbol{\hat{F}}_{\mathrm{ex}}$ is defined in Eq.~(\ref{eq:noise-operator}).

\textbf{Comparison to mean-field model.}
We compare Eqs.~(\ref{eq:coarse-graind_equation_dJ_dt}) with the existing mean-field theory and associate our model parameters with those obtained from experiments.
The mean-field spin operators are related to our formalism by $\langle\boldsymbol{\hat{\textbf{f}}}\rangle\equiv\int d^{3}r\langle\psi|\boldsymbol{\hat{\textbf{f}}}(\mathbf{r},t)|\psi\rangle/N_{\mathrm{a}}$ and $\langle\boldsymbol{\hat{\textbf{k}}}\rangle\equiv\int d^{3}r\langle\psi|\boldsymbol{\hat{\textbf{k}}}(\mathbf{r},t)|\psi\rangle/N_{\mathrm{b}}$, where $|\psi\rangle$ is the initial many-body wavefunction of the system.
Substituting these definitions in Eqs.~(\ref{eq:coarse-graind_equation_dJ_dt}) and using $\langle\boldsymbol{\hat{F}}_{\mathrm{ex}}\rangle=0$ recover the standard Bloch equations [Eqs.~\ref{eq:Bloch-dynamics}].

In terms of experimentally measured parameters, the interaction strength $\zeta$ is given by $\zeta=8\pi\kappa_{0}g_{\mathrm{e}}g_{\mathrm{n}}\mu_{\mathrm{B}}\mu_{\mathrm{n}}/(3q\hbar)$, where $g_{\mathrm{e}}=2$ is the electron g-factor, $g_{\mathrm{n}}$ is the g-factor of the noble-gas nucleus, $\mu_{\mathrm{B}}$ is Bohr magnetron, $\mu_{\mathrm{n}}$ is the magnetic moment of noble-gas spin, and $\kappa_{0}$ is the enhancement factor over the classical magnetic field due to the attraction of the alkali-metal electron to the noble-gas nucleus during a collision \cite{Happer2010book,Schaefer1989WalkerKappa0}.
For $\text{K\ensuremath{-^{3}}He}$ at $T=220^{\circ}\text{C}$, $\zeta=4.9\times10^{-15}\,\text{\ensuremath{\unit{cm^{3}}}}\text{s}^{-1}$ and $k_{\text{se}}=5.5\times10^{-20}\,\text{\ensuremath{\unit{cm^{3}}}}\text{s}^{-1}$ \cite{Happer2010book,Schaefer1989WalkerKappa0}.
Roughly estimating a collisional spin-exchange cross-section of  $\sigma\approx8\times10^{-15}\,\text{\ensuremath{\unit{cm^{2}}}}$ from the $\text{K\ensuremath{-^{3}}He}$ inter-atomic potential \cite{Pascale1983potentialHelium} and using a typical centrifugal potential with an angular momentum 40$\hbar$ yield an estimate of the precession angles $\langle\phi\rangle\approx\zeta q/\sigma v=1.4\times10^{-5}\,\text{rad}$ and $\langle\phi^{2}\rangle\approx4k_{\text{se}}/\sigma v=1.6\times10^{-10}\,\text{rad}^{2}$
for highly-polarized alkali vapor.

\textbf{Dynamics with diffusion and relaxation.}
To describe the spatial dynamics of a macroscopic ensemble of alkali and noble-gas spins in the presence of relaxation, we write the Heisenberg-Langevin equations for $\boldsymbol{\hat{\textbf{f}}}\left(\mathbf{r},t\right)$ and $\boldsymbol{\hat{\textbf{k}}}\left(\mathbf{r},t\right)$ 
\begin{equation}
\begin{aligned}\partial_{t}\boldsymbol{\hat{\textbf{f}}} & =-\frac{i}{\hbar}\bigl[\boldsymbol{\hat{\textbf{f}}},\mathcal{\mathcal{H}_{\mathrm{0}}+\mathcal{V}_{\mathrm{ex}}}\bigr]+D_{\mathrm{a}}\nabla^{2}\boldsymbol{\hat{\textbf{f}}}-\gamma_{\mathrm{a}}\boldsymbol{\hat{\textbf{f}}}+\boldsymbol{\hat{F}}_{\mathrm{a}}\\
\partial_{t}\boldsymbol{\hat{\textbf{k}}} & =-\frac{i}{\hbar}\bigl[\boldsymbol{\hat{\textbf{k}}},\mathcal{\mathcal{H}_{\mathrm{0}}+\mathcal{V}_{\mathrm{ex}}}\bigr]+D_{\mathrm{b}}\nabla^{2}\boldsymbol{\hat{\textbf{k}}}-\gamma_{\mathrm{b}}\boldsymbol{\hat{\textbf{k}}}+\boldsymbol{\hat{F}}_{\mathrm{b}}.
\end{aligned}
\label{eq:collective_dyn}
\end{equation}
Here we assume standard noble-gas pressures ($10^{-1}-10^{4}$ Torr), of which frequent collisions with the noble-gas atoms render the thermal motion diffusive, as described by the diffusion terms $D_{\mathrm{a}}\nabla^{2}\boldsymbol{\hat{\textbf{f}}}$ and $D_{\mathrm{b}}\nabla^{2}\boldsymbol{\hat{\textbf{k}}}$ \cite{Shaham2020Diffusion}.
For polarized alkali vapor, the interaction-free Hamiltonian from Eq.~(\ref{eq:H0_hamiltonian}) obtains the simple form $\mathcal{H}_{0}=\hbar\int\mathrm{d}^{3}\mathbf{r}[\tilde{\omega}_{\mathrm{a}}\hat{\text{f}}_{z}(\mathbf{r},t)+\tilde{\omega}_{\mathrm{b}}\hat{\text{k}}_{z}(\mathbf{r},t)]$, and $\mathcal{V}_{\mathrm{ex}}$ is given in Eq.~(\ref{eq:Coarse-grained-interaction-Hamiltonian}).
We emphasize that this model can describe the evolution of many-body quantum spin states.

Equations (\ref{eq:collective_dyn}) encapsulate the various spin dissipation mechanisms into the relaxation rates $\gamma_{\mathrm{a}}$ and $\gamma_{\mathrm{b}}$.
The relaxation rate of alkali-metal spins is given by 
\begin{equation}
\gamma_{\mathrm{a}}=n_{\mathrm{b}}(\sigma_{\text{sr}}v+k_{\text{se}})+n_{\mathrm{a}}\sigma_{\text{sd}}v_{\mathrm{a}}/2,
\end{equation}
consisting of collisional spin-orbit coupling, spin-exchange interaction with the noble-gas nuclei, and spin-destruction via binary collisions of alkali-metal spins (cross-section $\sigma_{\text{sd}}$, mean atomic velocity $v_\text{a}$) \cite{Happer2010book}.
The relaxation rate $\gamma$ in the main text can then be approximated by $\gamma=\gamma_\text{a}+D_{\mathrm{a}}\pi^{2}/R^{2}$, including the leading order of the diffusion relaxation in a spherical cell with radius $R$ (\textit{e.g.}, $R=1"$ in Fig.~\ref{fig:excitation exchange fidelity}) \cite{Shaham2020Diffusion}.
The relaxation rate of the noble-gas spins is given by 
\begin{equation}
\gamma_{\mathrm{b}}=k_{\text{se}}n_{\mathrm{a}}+T_{\mathrm{b}}^{-1},
\end{equation}
where $T_{\mathrm{b}}^{-1}$ is the coherence time in the absence of alkali atoms, usually limited by inhomogeneity of the magnetic field \cite{Walker2017He3review,Walker1997SEOPReview}.
We note that due to negligible noble-gas--cell-wall coupling, the diffusion-induced decay of the spatially-uniform mode of the noble gas vanishes \cite{Shaham2020Diffusion}.
The incoherent spin-transfer terms [third term in Eqs.~(\ref{eq:coarse-graind_equation_dJ_dt})], which have negligible effect on the coherent dynamics, are omitted for brevity.
In the experiments considered in the main text, $\gamma_{\mathrm{a}}^{-1}=50\,\mathrm{ms}$ and $\gamma_{\mathrm{b}}^{-1}=1-100$ hours, hence $\gamma_{\mathrm{b}}\lll\gamma_{\mathrm{a}}$.
The Langevin noise operators $\boldsymbol{\hat{F}}_{\mathrm{a}}$ and $\boldsymbol{\hat{F}}_{\mathrm{b}}$ in Eq.~(\ref{eq:collective_dyn}) account for fluctuations and for preserving commutation relations under the relaxations $\gamma_{\mathrm{a}}$,$\gamma_{\mathrm{b}}$ and diffusion \cite{GardinerZoller2004quantumNoise}; their manifestation as scalar operators $\hat{F}_{\mathrm{a}}$ and $\hat{F}_{\mathrm{b}}$ after the Holstein-Primakoff approximation is given below.

As described in the main text, we consider highly-polarized ensembles with most spins pointing downward ($-\hat{z}$) \cite{Julsgaard2001PolzikEntanglement,Polzik2010ReviewRMP,Sherson2006PolzikTeleportationDemo}, approximate $\text{f}_{z}=-p_{\mathrm{a}}n_{\mathrm{a}}q/2$ and $\mathrm{k}_{z}=-p_{\mathrm{b}}n_{\mathrm{b}}/2$, and apply the Holstein-Primakoff transformation \cite{Polzik2010ReviewRMP} to represent the collective states as excitations of a bosonic field with local annihilation operators $\hat{a}\left(\mathbf{r},t\right)=\hat{\text{f}}_{-}\left(\mathbf{r},t\right)/\sqrt{2|\text{f}_{z}|}$ and $\hat{b}\left(\mathbf{r},t\right)=\hat{\text{k}}_{-}\left(\mathbf{r},t\right)/\sqrt{2|\text{k}_{z}|}$.
The creation operators $\hat{a}^{\dagger}\left(\mathbf{r},t\right)$ and $\hat{b}^{\dagger}\left(\mathbf{r},t\right)$ flip upwards one alkali or noble-gas spin at position $\mathbf{r}$.

When the two gases are polarized, the energy cost of flipping a spin in one species is the sum of Zeeman shift (due to the external magnetic field) and so-called collisional shift (due to effective magnetic field induced by the other species)  \cite{Schaefer1989WalkerKappa0}.
The altered Larmor frequencies $\omega_{\mathrm{a}}=\tilde{\omega}_{\mathrm{a}}-\zeta p_{\mathrm{b}}n_{\mathrm{b}}/2$ and $\omega_{\mathrm{b}}=\tilde{\omega}_{\mathrm{b}}-\zeta p_{\mathrm{a}}n_{\mathrm{a}}q/2$ are obtained when rewriting Eqs.~(\ref{eq:collective_dyn}) in terms of $\hat{a}(\mathbf{r},t)$ and $\hat{b}(\mathbf{r},t),$ 
\begin{equation}
\begin{aligned}\partial_{t}\hat{a} & =-\bigl(i\omega_{\mathrm{a}}+\gamma_{\mathrm{a}}-D_{\mathrm{a}}\nabla^{2}\bigr)\hat{a}-iJ\hat{b}+\hat{F}_{\mathrm{a}},\\
\partial_{t}\hat{b} & =-\bigl(i\omega_{\mathrm{b}}+\gamma_{\mathrm{b}}-D_{\mathrm{b}}\nabla^{2}\bigr)\hat{b}-iJ\hat{a}+\hat{F}_{\mathrm{b}}.
\end{aligned}
\label{eq:collective_dyn-1}
\end{equation}
Importantly, here we obtain the coherent spin-exchange rate $J=\zeta\sqrt{qp_{\mathrm{a}}p_{\mathrm{b}}n_{\mathrm{a}}n_{\mathrm{b}}}/2$, responsible for the local coupling of the two collective spins, as illustrated in Fig.~\ref{fig:illustration}c.

\textbf{Numerical solution.}
We numerically solve the differential Eqs.~(\ref{eq:collective_dyn-1}) for two particular cases, using the experimental parameters outlined in the main text.
The analytical solution of Eqs.~(\ref{eq:collective_dyn-1}) utilizes a decomposition to coupled spatial (diffusion) modes, given in the SI.
Our calculation uses the first 100 spherically symmetric least-decaying modes.
We assume an alkali polarization of $p_{\text{a}}=0.95$ and noble-gas spin polarization of $p_{\text{b}}=0.75$, corresponding to initial $\langle\hat{a}^{\dagger}\hat{a}\rangle=0.05$ and $\langle\hat{b}^{\dagger}\hat{b}\rangle=0.17$ at $t=0$.
We account for these incoherent excitations, due to the imperfect polarizations, by using a non-pure density-matrix and incorporating the associated (excess) quantum noise.

First, the spatially-symmetric (uniform) mode of the alkali spin ensemble is initialized in a Fock state with two spin excitations.
The probability of transferring the excitations to the uniform mode of the noble-gas spins is presented in Fig.$\,$\ref{fig:excitation exchange fidelity}a.
Here, by changing $\Delta$ at $t=\pi/(2J)$, the excitation transfer is complete, and the noble gas becomes decoupled from the alkali and free from collision-induced relaxation.
Second, the uniform mode of the alkali spin ensemble is initialized in a squeezed vacuum state with $\text{7\,dB }$of squeezing.
The squeezing is transferred to the noble-gas spins with high fidelity, as presented in Fig.$\,$\ref{fig:excitation exchange fidelity}b.
We include 100 modes to ensure convergence of the calculations, with no significant improvement observed with additional modes.

\section*{Noise associated with spin-exchange collisions}

The fluctuation vector-operator $\boldsymbol{\hat{F}}_{\mathrm{ex}}$ in Eqs.~(\ref{eq:coarse-graind_equation_dJ_dt}) can be defined by (cf. Ref.~\cite{Gabrielli2014VlasovEQN})
\begin{align}
\boldsymbol{\hat{F}}_{\mathrm{ex}}(\mathbf{r},t) & \mathrm{d}t=-\zeta\boldsymbol{\hat{\textbf{k}}}(\mathbf{r},t)\times\boldsymbol{\hat{\textbf{f}}}(\mathbf{r},t)dt\label{eq:noise-operator}\\
+ & \frac{1}{\tau'}\int_{t}^{t+dt}ds\sum_{ab}\varkappa_{ab}(s,\tau')\phi_{a}(s)w(\mathbf{r}_{b}-\mathbf{r}_{a})\boldsymbol{\hat{\textbf{k}}}_{b}\times\boldsymbol{\hat{\textbf{s}}}_{a},\nonumber 
\end{align}
accounting for the stochastic fine-grained dynamics [of Eq.~(\ref{eq:spin-operators-discrete-step-evolution1})] within the coarse-grained description [of Eq.~(\ref{eq:coarse-graind_equation_dJ_dt})].
Similarly to what we observe in the toy-model simulation (Fig.~\ref{fig:stochastic-simulation} and SI), the operator $\boldsymbol{\hat{F}}_{\mathrm{ex}}$ describes fluctuations of order $\langle\phi^{2}\rangle$ in the coherent mutual-precession process, manifesting a stochastic superposition of non-symmetric local spin-operators.
Fluctuation in the incoherent terms are of order $\langle\phi^{4}\rangle$ and thus negligible.

We now examine the statistical properties of the fluctuation operator $\boldsymbol{\hat{F}}_{\mathrm{ex}}$.
It is zero on average $\bigl\langle\boldsymbol{\hat{F}}_{\mathrm{ex}}(\mathbf{r},t)\bigr\rangle=0,$ and its correlations satisfy 
\begin{align*}
\bigl\langle\hat{F}_{\mathrm{ex},i}(\mathbf{r},t)\hat{F}_{\mathrm{ex},j}(\mathbf{r}',t')\bigr\rangle & =\frac{1}{2}n_{\mathrm{a}}n_{\mathrm{b}}k_{\text{se}}\delta\left(t-t'\right)w(\mathbf{r}-\mathbf{r}')\hat{L}_{ij},
\end{align*}
where $\hat{L}_{ij}=\delta_{ij}\mathds{1}-2\{\hat{\text{k}}_{i},\hat{\text{s}}_{j}\}/(n_{\mathrm{a}}n_{\mathrm{b}})+i\epsilon_{ijm}\bigl(\hat{\text{k}}_{m}/n_{\mathrm{b}}+\hat{\text{s}}_{m}/n_{\mathrm{a}}\bigr)$.
Here $i,j,m\in\left\{ x,y,z\right\} $ and the symbol $\epsilon_{ijm}$ is the Levi-Civita tensor.
We interpret $\boldsymbol{\hat{F}}_{\mathrm{ex}}$ as temporally and spatially white, since its correlations are proportional to $\delta\left(t-t'\right)$ and to the coarse-grained delta function $w(\mathbf{r}-\mathbf{r}')$.
Furthermore, the coarse-grained commutation relations $\bigl[\hat{\text{s}}_{i}(\mathbf{r},t),\hat{\text{s}}_{j}(\mathbf{r}',t)\bigr]=iw(\mathbf{r}-\mathbf{r}')\epsilon_{ijm}\hat{\text{s}}_{m}(\mathbf{r},t)$ and $\bigl[\hat{\text{k}}_{i}(\mathbf{r},t),\hat{\text{k}}_{j}(\mathbf{r}',t)\bigr]=iw(\mathbf{r}-\mathbf{r}')\epsilon_{ijm}\hat{\text{k}}_{m}(\mathbf{r},t)$ are preserved.
Indeed, the relaxation of the commutation relations after $dt$ due to the loss terms in Eqs.~(\ref{eq:coarse-graind_equation_dJ_dt}) is exactly balanced by the fluctuations $\hat{F}_{\mathrm{ex},i}(\mathbf{r},t)\hat{F}_{\mathrm{ex},j}(\mathbf{r}',t)dt^{2}$.
We therefore formally identify $\boldsymbol{\hat{F}}_{\mathrm{ex}}$ as a quantum white noise operator \cite{GardinerZoller2004quantumNoise}
originating from the randomness of the collisional interaction.

For fully-polarized spin ensembles, the corresponding noise correlations are found to have the standard form of a vacuum noise \cite{GardinerZoller2004quantumNoise}, satisfying
\begin{equation}
\begin{aligned}\bigl\langle\hat{F}_{\mathrm{ex}}^{-}(\mathbf{r},t)\hat{F}_{\mathrm{ex}}^{+}(\mathbf{r}',t')\bigr\rangle & =2n_{\mathrm{a}}n_{\mathrm{b}}k_{\text{se}}\delta\left(t-t'\right)w(\mathbf{r}-\mathbf{r}'),\\
\bigl\langle\hat{F}_{\mathrm{ex}}^{+}(\mathbf{r},t)\hat{F}_{\mathrm{ex}}^{-}(\mathbf{r}',t')\bigr\rangle & =0.
\end{aligned}
\end{equation}
The contribution of $\boldsymbol{\hat{F}}_{\mathrm{ex}}$ to the alkali noise operator $\hat{F}_{\mathrm{a}}$ in Eqs.~(\ref{eq:coherent-dynamics}) and (\ref{eq:collective_dyn-1}) is given by $\hat{F}_{\mathrm{a}}=\hat{F}_{\mathrm{ex}}^{-}/\sqrt{2|\text{f}_{z}|}$.
Therefore, the variance is $\bigl\langle\hat{F}_{\mathrm{a}}(\mathbf{r},t)\hat{F}_{\mathrm{a}}^{\dagger}(\mathbf{r}',t')\bigr\rangle=2\gamma_{\text{ex}}\delta\left(t-t'\right)w(\mathbf{r}-\mathbf{r}'),$ where $\gamma_{\text{ex}}=n_{\mathrm{b}}k_{\text{se}}/q$ is the spin-exchange relaxation for the alkali spins.
We thus conclude that, for polarized ensembles, the spin-exchange noise appears as a vacuum noise, which is the minimal possible noise.

For general spin polarizations $p_\text{a},p_\text{b}\le 1$, the second-order moments of the noise are given by 
\begin{align*}
\bigl\langle\hat{F}_{\mathrm{a}}(\mathbf{r},t)\hat{F}_{\mathrm{a}}^{\dagger}(\mathbf{r}',t')\bigr\rangle & =\frac{2+p_{\text{a}}+p_{\text{b}}}{4p_{\text{a}}}2\gamma_{\text{ex}}\delta\left(t-t'\right)w(\mathbf{r}-\mathbf{r}')\\
\bigl\langle\hat{F}_{\mathrm{a}}^{\dagger}(\mathbf{r},t)\hat{F}_{\mathrm{a}}(\mathbf{r}',t')\bigr\rangle & =\frac{2-p_{\text{a}}-p_{\text{b}}}{4p_{\text{a}}}2\gamma_{\text{ex}}\delta\left(t-t'\right)w(\mathbf{r}-\mathbf{r}').
\end{align*}
Importantly, for highly-polarized ensembles ($1-p_\text{a}\ll1$, $1-p_\text{b}\ll1$), the excess noise is small, since  $\frac{2-p_{\text{a}}-p_{\text{b}}}{4p_{\text{a}}}\ll 1$ and $\frac{2+p_{\text{a}}+p_{\text{b}}}{4p_{\text{a}}}-1\ll 1$.
It is interesting to note that imperfect polarization contributes to other quantum noise terms as well, and this contribution is quantitatively similar to that in $\boldsymbol{\hat{F}}_{\mathrm{ex}}$ and $\hat{F}_{\mathrm{a}}$.
It follows that the coherent interface based on spin-exchange collisions is not much more sensitive to imperfect polarization compared to other quantum effects in single-specie spin systems.

In practice, the relaxation due to spin exchange, including the effect of $\boldsymbol{\hat{F}}_{\mathrm{ex}}$, is small compared to that originating from other sources.
For example, the relaxation rate of the alkali electron spin due to spin exchange is $n_{\mathrm{b}}k_{\text{se}}$, whereas the relaxation rate due the spin-orbit coupling during collisions is $n_{\mathrm{b}}\sigma_{\text{sr}}v$, where $\sigma_{\text{sr}}$ is the spin-rotation cross-section.
The relative importance of these two mechanisms is characterized by the parameter $\eta=k_{\text{se}}/(\sigma_{\text{sr}}v)$, where $\eta=0.34$ for potassium-helium, $\eta=0.024$
for rubidium-helium, and $\eta<0.01$ for
cesium-helium at $215~^{\circ}\text{C}$
\cite{Happer2010book}.
The relaxation of the noble-gas spins due to spin exchange with alkali spins is negligible for $t\ll(n_{\mathrm{a}}k_{\text{se}})^{-1} \approx 17\,\text{hours}$ when operating with $n_{\mathrm{a}}=3\times10^{14}\,\text{\ensuremath{\unit{cm^{-3}}}}$.
When perfectly-polarized ensembles are initialized, quantum noises other than spin exchange also behave as vacuum noises.
In this case, the noise terms $\hat{F}_{\mathrm{a}}=(\hat{F}_{\mathrm{a},x}+i\hat{F}_{\mathrm{a},y})/\sqrt{2|\text{f}_{z}|}$ and $\hat{F}_{\mathrm{b}}=(\hat{F}_{\mathrm{b},x}+i\hat{F}_{\mathrm{b},y})/\sqrt{2|\text{k}_{z}|}$ satisfy $\langle\hat{F}_{\chi}\rangle=\langle\hat{F}_{\chi}^{\dagger}\hat{F}_{\chi}\rangle=0$ and $[\hat{F}_{\chi}(\mathbf{r},t),\hat{F}_{\chi}^{\dagger}(\mathbf{r}',t')]=\langle\hat{F}_{\chi}(\mathbf{r},t)\hat{F}_{\chi}^{\dagger}(\mathbf{r}',t')\rangle=[2\gamma_{\chi}w(\mathbf{r}-\mathbf{r}')+C_{\chi}(\mathbf{r},\mathbf{r}')]\delta\left(t-t'\right)$ for $\chi\in\left\{ \mathrm{a,b}\right\} $, including the fluctuations induced by the spin-exchange interaction.
The function $C_{\chi}(\mathbf{r},\mathbf{r}')$ is the diffusion component of the noise correlation function \cite{Shaham2020Diffusion}, independent of the spin-exchange interaction or of the other relaxation mechanisms incorporated in $\gamma_{\chi}$.

Finally, the incoherent excitations due to imperfect polarization also add a classical (mix state) background to the (otherwise pure) expectation values, which reduces the significance or fidelity of non-classical phenomena.
For example, the probability of finding a collective excitation in an otherwise unexcited ensemble is $2p_\chi(1-p_\chi)/(1+p_\chi)^2$.

\begin{acknowledgments}
We thank C. Avinadav for helpful discussions.
We acknowledge financial support by the Israel Science Foundation and ICORE, the European Research Council starting investigator grant Q-PHOTONICS 678674, the Pazy Foundation, the Minerva Foundation with funding from the Federal German Ministry for Education and Research, and the Laboratory in Memory of Leon and Blacky Broder.
\end{acknowledgments}

\bibliography{weak_collisions}

\onecolumngrid \appendix 

\newpage

\setcounter{equation}{0}
\setcounter{figure}{0}
\setcounter{table}{0}
\setcounter{page}{1}
\makeatletter
\renewcommand{\theequation}{S\arabic{equation}}
\renewcommand{\thefigure}{S\arabic{figure}}
\renewcommand{\bibnumfmt}[1]{[S#1]}
\renewcommand{\citenumfont}[1]{S#1} 

\part*{\centerline{Supplementary Information}}

\subsection{Stochastic simulation of spin-exchange collisions}

Here we detail the mathematical formulation of the stochastic numerical simulations used in Fig.~\ref{fig:stochastic-simulation} in the main text and the derivation of the properties of a single spin excitation in Eq.~(\ref{eq:spin-detail}) in the main text.
Our simulations solve the Hamiltonian time-evolution of a quantum state of $N_{\mathrm{a}}$ electron spins (alkali-like atoms with $S=1/2$) and $N_{\mathrm{b}}$ noble-gas spins ($K=1/2$) when nearly all spins point downwards.
In each simulation time-step, we pair each electron spin (labeled $a$) with a single random noble-gas spin (labeled $b$).
The pair selections at time-step $n$ is encapsulated in the random variable $\varkappa_{ab}^{(n)}$, with $\varkappa_{ab}^{(n)}=1$ if spin $a$ and $b$ collide and $\varkappa_{ab}^{(n)}=0$ otherwise.
We constrain the pairing process by $\varkappa_{ab}^{(n)}\varkappa_{cb}^{(n)}=0$ for $a\neq c$, ensuring that each noble-gas spin interacts at most with a single electron spin at each time step.
The collision between spins $a$ and $b$ is described by the exchange Hamiltonian 
\[
\mathcal{V}_{ab}=\frac{\phi_{ab}}{\tau}\bigl(\hat{\mathrm{s}}_{az}\hat{\mathrm{k}}_{bz}+\frac{1}{2}\hat{\mathrm{s}}_{a+}\hat{\mathrm{k}}_{b-}+\frac{1}{2}\hat{\mathrm{k}}_{b+}\hat{\mathrm{s}}_{a-}\bigr),
\]
where $\hat{\mathrm{s}}_{a\pm}$ and $\hat{\mathrm{k}}_{b\pm}$ are ladder spin operators.
We also include a magnetic interaction for the alkali spins via the Hamiltonian term $\gamma_{\mathrm{a}}B_{z}\sum_{a}\hat{\mathrm{s}}_{az}$.
We set $\gamma_{\mathrm{a}}B_{z}\tau=\langle\phi\rangle/2$ to simulate $\Delta=0$ and enter the strong coupling regime.
The scattering-matrix per electron spin $a$ at time-step $n$ is given by
\begin{equation}
\begin{aligned}U_{n}^{(a)} & =\sum_{b}\varkappa_{ab}^{(n)}e^{\tfrac{i}{2}\phi_{a}^{(n)}}\cdot\left[i\sin(\tfrac{1}{2}\phi_{a}^{(n)})\cdot\left(\left|\downarrow\uparrow\right\rangle \left\langle \uparrow\downarrow\right|+\left|\uparrow\downarrow\right\rangle \left\langle \downarrow\uparrow\right|\right)\right.\\
- & \left.2\sin^{2}(\tfrac{1}{4}\phi_{a}^{(n)})\cdot\left(\left|\uparrow\downarrow\right\rangle \left\langle \uparrow\downarrow\right|+\left|\downarrow\uparrow\right\rangle \left\langle \downarrow\uparrow\right|\right)\right]+\mathds{1},
\end{aligned}
\label{a}
\end{equation}
where $\left|\downarrow\uparrow\rangle\right.\equiv\left|\downarrow_{a}\rangle_{\mathrm{a}}\right.\left|\uparrow_{b}\rangle_{\mathrm{b}}\right.$, $\left|\uparrow\downarrow\rangle\right.\equiv\left|\uparrow_{a}\rangle_{\mathrm{a}}\right.\left|\downarrow_{b}\rangle_{\mathrm{b}}\right.$, and $\mathds{1}$ is the identity operator.
This form describes the dynamics at the compensation point, where the precession of the two gases is synchronized $(\omega_{\mathrm{a}}=\omega_{\mathrm{b}})$.
The collision angle $\phi_{a}^{(n)}$ is a random variable, sampled from a Gaussian distribution $\mathcal{N}(\langle\phi\rangle,\delta\phi)$ with mean $\langle\phi\rangle$ and typical standard deviation of $\delta\phi=\langle\phi\rangle$, where $\langle\phi\rangle\ll1$ is an input parameter.
The wavefunction of the system then evolves by $\left|\psi(n+1)\rangle\right.=U_{n}|\psi(n)\rangle$, where the time-evolution operator $U_{n}$ is given by $U_{n}=\Pi_{a}U_{n}^{(a)}$.
The duration of each time step is the mean time between collisions of an alkali atom with any of the noble-gas atoms $\tau$.
In units of $\tau$, time is defined as $t=n$.
To simulate the physical apparatus presented in the main text, we consider here $\tau=(n_{\mathrm{b}}\sigma v_{\mathrm{T}})^{-1}$, where $\sigma$ is the hard-sphere collisions cross-section and $v_{\mathrm{T}}$ is the colliding atoms reduced velocity, being the mean time between subsequent collisions of a given alkali atom with various noble gas atoms.

The simulation results presented in Figs.~\ref{fig:stochastic-simulation}(a-d) are restricted to the Hilbert subspace of a single spin-up excitation, represented by the wavefunctions $|\psi(t)\rangle=\sum_{i=1}^{N_{\mathrm{a}}+N_{\mathrm{b}}}c_{i}(t)|\downarrow\ldots\downarrow\uparrow_{i}\downarrow\ldots\downarrow\rangle_{\mathrm{a\&b}}$ for a set of complex-valued coefficients $c_{i}(t)$ satisfying $\sum_{i}|c_{i}(t)|^{2}=1$.
Note that this subspace is invariant to the exchange operation due to conservation of the total spin.
We simulate the system for $N_{\mathrm{a}}=100$, $N_{\mathrm{b}}=10^{4},$ $\langle\phi\rangle=10^{-5}$, with either initial symmetric excitations $|\psi_{0}\rangle=|1\rangle_{\mathrm{a}}|0\rangle_{\mathrm{b}}$ (Fig.~\ref{fig:stochastic-simulation}a) or localized excitations $|\psi_{0}\rangle=|\uparrow\downarrow\ldots\downarrow\rangle_{\mathrm{a}}|0\rangle_{\mathrm{b}}$ (Fig.~\ref{fig:stochastic-simulation}b).
The contrast in the latter case scales with the small overlap $\langle1|_{\mathrm{a}}|\uparrow\downarrow\ldots\downarrow\rangle_{\mathrm{a}}=1/\sqrt{N_{\mathrm{a}}}$.
The rate of the collective oscillations is found to be $2J=\sqrt{N_{\mathrm{a}}/N_{\mathrm{b}}}\langle\phi\rangle/\tau$, fully agreeing with the mean field and many-body analysis appearing in the main text.

In Figs.~\ref{fig:stochastic-simulation}c and \ref{fig:stochastic-simulation}d, we set $\langle\phi\rangle=2\cdot10^{-2}$ and repeat the simulations.
We observe the electron spin dephasing and thermalization with the noble-gas spins at a rate $\gamma_\text{a}=\langle\phi^{2}\rangle/(4\tau)$ (Fig.~\ref{fig:stochastic-simulation}d).
This decay is reduced when the two spin ensembles are strongly coupled and the excitation is also shared by the noble gas spins (Fig.~\ref{fig:stochastic-simulation}c).
We also verified that exchange and decoherence weakly depend on the chosen distribution parameters.
The exchange rate $J$ always satisfies $2J=\sqrt{N_{\mathrm{a}}/N_{\mathrm{b}}}\langle\phi\rangle/\tau$, while the decoherence rate (in terms of the decay of fidelity) satisfies $\gamma_\text{a}=\alpha\langle\phi^{2}\rangle/(4\tau)$ with $\alpha\le1$.
For fully polarized noble-gas spins $\alpha=1/2$ when $\delta\phi\ll\langle\phi\rangle$ and the randomness of the collisions process results solely from randomness in pairing of colliding atoms.
When $\delta\phi\gg\langle\phi\rangle$, $\alpha=1$ due to additional randomness in collisions intensity, making collisions less correlated.
These values fully agree with mean-field and many-body analytic solutions.
Note also that the noble-gas spin dephasing $\gamma_{\mathrm{b}}=N_{\mathrm{a}}/N_{\mathrm{b}}\cdot\gamma_{\mathrm{a}}$ is negligible since $N_{\mathrm{a}}\ll N_{\mathrm{b}}$.
In Fig.~\ref{fig:stochastic-simulation}e, we simulate the system evolution in the spin-exchange-invariant subspace comprising two spin excitations $|\psi(t)\rangle=\sum_{i\ne j}^{N_{\mathrm{a}}+N_{\mathrm{b}}}c_{ij}(t)\left|\downarrow\ldots\downarrow\uparrow_{i}\downarrow\ldots\downarrow\uparrow_{j}\downarrow\ldots\downarrow\right\rangle _{\mathrm{a\&b}}$ for $\sum_{i\ne j}|c_{ij}(t)|^{2}=1$.
Here we use $N_{\mathrm{a}}=30$, $N_{\mathrm{b}}=300$, $\langle\phi\rangle=10^{-5}$, and $|\psi_{0}\rangle=|1\rangle_{\mathrm{a}}|1\rangle_{\mathrm{b}}$.
These simulations demonstrate an analogy to the nonclassical Hong-Ou-Mandel phenomenon utilizing a beam-splitter with variable reflectivity.
At $t=\pi m/(2J)$ for any integer $m$, the two excitations are bunched in either of the two spin ensembles (analogous to the ports of the beam-splitter), generating an entangled state.

For a system initialized with a symmetric excitation $|\psi_{0}\rangle=|1\rangle_{\mathrm{a}}|0\rangle_{\mathrm{b}}$, for short timescales and to first order in $\phi$ the evolution is given by 
\[
|\psi(t/\tau)\rangle=\bigl[\mathds{1}-i\sum_{n}^{t/\tau}\sum_{ab}\varkappa_{ab}^{(n)}\phi_{a}^{(n)}\bigl(\frac{1}{2}\hat{\mathrm{s}}_{a-}\hat{\mathrm{k}}_{b+}+\hat{\mathrm{s}}_{z}\hat{\mathrm{k}}_{z}\bigr)\bigr]|\psi_{0}\rangle,
\]
using $\hat{\mathrm{s}}_{a+}\hat{\mathrm{k}}_{b-}|\psi_{0}\rangle=0$ (since no excitations populate the initial state of the noble-gas).
The second term is then decomposed into the deterministic and stochastic terms, $-iJt|1\rangle_{\mathrm{a}}|0\rangle_{\mathrm{b}}$ and $-i\epsilon|\delta\psi\rangle$, as outlined in the main text.
Here we identify the stochastic transition amplitude 
\[
\epsilon=[\sum_{b}(\sum_{a}\sum_{n=0}^{t/\tau}\delta\phi_{ab}^{(n)}){}^{2}/(2N_{\mathrm{a}})]{}^{1/2}
\]
and the stochastic wavefunction $|\delta\psi\rangle\equiv|0\rangle_{\mathrm{a}}|\delta\psi\rangle_{\mathrm{b}}+|\delta\psi\rangle_{\mathrm{a}}|0\rangle_{\mathrm{b}}$ where 
\[
\epsilon|\delta\psi\rangle_{\mathrm{b}}=\tfrac{1}{2\sqrt{N_{\mathrm{a}}}}\sum_{b}\bigl[\sum_{an}\delta\phi_{ab}^{(n)}\bigr]\hat{\mathrm{k}}_{b+}|0\rangle_{\mathrm{b}}
\]
describes stochastic non-symmetric noble-gas spin excitations while
\[
\epsilon|\delta\psi\rangle_{\mathrm{a}}=\tfrac{1}{2\sqrt{N_{\mathrm{a}}}}\sum_{a}\bigl[\sum_{bn}\delta\phi_{ab}^{(n)}\bigr]\hat{\mathrm{s}}_{a+}|0\rangle_{\mathrm{a}}.
\]
describes non-symmetric excitations of alkali spins.
These wavefunctions depend on the fluctuation of the spins precession $\delta\phi_{ab}^{(n)}=\varkappa_{ab}^{(n)}\phi_{a}^{(n)}-\langle\phi\rangle/N_{\mathrm{b}}$ where $\langle\phi\rangle\equiv\langle\phi_{a}^{(n)}\rangle$.
Following the central limit and assuming the stringent condition that any two collisions are uncorrelated $\langle\delta\phi_{a'b}^{(n')}\delta\phi_{ab}^{(n)}\rangle=\delta_{aa'}\delta_{bb'}\delta_{nn'}\langle(\delta\phi_{ab}^{(n)})^{2}\rangle$, we find the result in the main text $\epsilon\rightarrow\sqrt{\langle\phi^{2}\rangle t/(2\tau)}$ where $\langle\phi^{2}\rangle\equiv\langle[\phi_{a}^{(n)}]^{2}\rangle$.

\subsection{Coupled spatial modes\label{subsec:Coupled-spatial-modes}}

Here we present a numerical solution for Eq.~(\ref{eq:collective_dyn-1}) in the Methods section, using a set of mode-function basis.
This decomposition is illustratively shown in Fig.~\ref{fig:illustration}d and was used in the numerical calculation in Fig.~\ref{fig:excitation exchange fidelity}.

The irreversibility of the evolution in Eq.~(\ref{eq:collective_dyn-1}) is dominated by alkali-spin relaxation and by spatial atomic diffusion.
In principle, absence the diffusion, the dynamics of $\hat{a}\left(\mathbf{r},t\right)$ and $\hat{b}\left(\mathbf{r},t\right)$ at any location $\mathbf{r}$ would be unitary and deterministic for short times $t\ll(\gamma_{\mathrm{a}}+\gamma_{\mathrm{b}})^{-1}$.
That could allow for local (multi-mode) coupling of the two gases, owing to the locality of the collisional interaction.
In practice, however, the diffusion term is dominant in the dynamics \cite{Firstenberg2013RMPcoherentDiffusion}.
It is therefore fruitful to consider the spatial modes $\hat{a}_{m}\left(t\right)\equiv\int A_{m}\left(\mathbf{r}\right)\hat{a}\left(\mathbf{r},t\right)\mathrm{d}^{3}\mathbf{r}$ and $\hat{b}_{n}\left(t\right)\equiv\int B_{n}\left(\mathbf{r}\right)\hat{b}\left(\mathbf{r},t\right)\mathrm{d}^{3}\mathbf{r}$, associated with orthonormal and complete sets of eigenmodes $A_{m}\left(\mathbf{r}\right)$ and $B_{n}\left(\mathbf{r}\right)$ of the respective diffusion-relaxation operators $D_{\mathrm{a}}\nabla^{2}-\gamma_{\mathrm{a}}$ and $D_{\mathrm{b}}\nabla^{2}-\gamma_{\mathrm{b}}$.
Typically one could assume partial or full relaxation of alkali spins by the cell walls (Robin boundary conditions) and no relaxation of noble-gas spins at the walls.
The corresponding eigenvalues $\gamma_{\mathrm{a}m}$ and $\gamma_{\mathrm{b}n}$ associate a decay rate with each mode.

The dynamics of the spatial modes $\hat{a}_{m}$ and $\hat{b}_{n}$, illustrated in Fig.~\ref{fig:illustration}d, is described by coupled-mode equations 
\begin{equation}
\begin{array}{cc}
\partial_{t}\hat{a}_{m} & =-(i\omega_{\mathrm{a}}+\gamma_{\mathrm{a}m})\hat{a}_{m}-iJ\sum_{n}c_{mn}\hat{b}_{n}+\hat{F}_{\textnormal{a}m}\\
\partial_{t}\hat{b}_{n} & =-(i\omega_{\mathrm{b}}+\gamma_{\mathrm{b}n})\hat{b}_{n}-iJ\sum_{m}c_{mn}^{*}\hat{a}_{m}+\hat{F}_{\textnormal{b}n}
\end{array}\label{eq:mode_equation_a}
\end{equation}
with an effective coupling $Jc_{mn}$ that is determined by their integrated spatial overlap coefficients $c_{mn}=\int A_{m}\left(\mathbf{r}\right)B_{n}^{*}\left(\mathbf{r}\right)\mathrm{d}^{3}\mathbf{r}$, being elements of a unitary matrix.
The noise terms of the $m$-th and $n$-th modes are $\hat{F}_{\textnormal{a}m}=\int A_{m}\left(\mathbf{r}\right)\hat{F}_{\textnormal{a}}\left(\mathbf{r}\right)\mathrm{d}^{3}\mathbf{r}$ and $\hat{F}_{\textnormal{b}n}=\int B_{n}\left(\mathbf{r}\right)\hat{F}_{\textnormal{b}}\left(\mathbf{r}\right)\mathrm{d}^{3}\mathbf{r}$.

Eqs.~(\ref{eq:mode_equation_a}) describe the coupling of any quantum mode of one gas with the $N_{\text{modes}}\approx V/V_l$ modes of the other gas.
In practice however, the majority of modes are barely coupled.
It is constructive to differentiate between the set of high-order modes of the diffusion operator, defined by $\mathcal{R}\equiv\bigl\{\hat{a}_{r},\hat{b}_{r}|\gamma_{\mathrm{a}r},\gamma_{\mathrm{b}r}\gg J\bigr\}$ ($0\leq r<N_{\text{modes}}$) and the complementary set of stable modes $\mathcal{S}=1\setminus\mathcal{R}$.
The high-order modes $\mathcal{R}$ are characterized by rapid relaxation due to the thermal motion of the atoms.
These modes experience little coherent interaction and, at long timescales $dt\gg1/\gamma_{\mathrm{a}r},1/\gamma_{\mathrm{b}r}$, are governed by 
\begin{equation}
\begin{array}{cc}
\hat{a}_{r}(t) & =\hat{w}_{\mathrm{a}r}(t)-\frac{iJ}{\gamma_{\mathrm{a}r}-i\Delta}\sum_{n\in\text{S}}c_{rn}\hat{b}_{n}(t)\\
\hat{b}_{r}(t) & =\hat{w}_{\mathrm{b}r}(t)-\frac{iJ}{\gamma_{\mathrm{b}r}+i\Delta}\sum_{m\in\text{S}}c_{mr}^{*}\hat{a}_{m}(t).
\end{array}\label{eq:reservoir_mode_a_rt}
\end{equation}
The first terms, $\hat{w}_{\mathrm{q}r}(t)=\int_{0}^{t}d\tau'e^{-(i\omega_{\mathrm{q}}+\gamma_{\mathrm{q}r})\tau'}\hat{F}_{\mathrm{q}r}(t-\tau')$ for $q\in\left\{ \mathrm{a},\mathrm{b}\right\} $, describe the diffusion-induced quantum process.
This process dominates the dynamics, which is Markovian: any dependence on $\hat{a}_{r}(t_{0})$ and $\hat{b}_{r}(t_{0})$
at $t_{0}\ll t$ is erased exponentially at the fast rates $\gamma_{\mathrm{a}r}$
and $\gamma_{\mathrm{b}r}$.
Consequently, the modes $\hat{a}_{r}(t)$ and $\hat{b}_{r}(t)$ can be considered as thermal reservoirs.
The second term in Eqs.~(\ref{eq:reservoir_mode_a_rt}) describes the weak coupling to the stable modes $\hat{a}_{s},\hat{b}_{s}\in\mathcal{S}$ via the coherent collisional interaction.
Substitution of Eqs.~(\ref{eq:reservoir_mode_a_rt}) in Eqs.~(\ref{eq:mode_equation_a}) yields a relatively small and close set of coupled equations for the stable modes, governed by a coherent dynamics 
\begin{equation}
\begin{aligned}\partial_{t}\hat{a}_{s} & =-(i\omega_{\mathrm{a}}+\gamma_{\mathrm{a}s})\hat{a}_{s}-J\sum_{n\in\text{S}}(ic_{sn}\hat{b}_{n}+\epsilon_{sn}^{(\mathrm{a})}\hat{a}_{n})+\hat{G}_{\mathrm{a}s},\\
\partial_{t}\hat{b}_{s} & =-(i\omega_{\mathrm{b}}+\gamma_{\mathrm{b}s})\hat{b}_{s}-J\sum_{m\in\text{S}}\bigl(ic_{ms}^{*}\hat{a}_{m}+\epsilon{}_{sm}^{(\mathrm{b})*}\hat{b}_{m}\bigr)+\hat{G}_{\mathrm{b}s},
\end{aligned}
\label{eq:stable-mode-a-EOM}
\end{equation}
The coefficients $\epsilon_{sn}^{(\mathrm{a})}=\sum_{r\in R}c_{sr}c_{nr}^{*}J/(\gamma_{\mathrm{b}r}+i\Delta)$ and $\epsilon_{sn}^{(\mathrm{b})}=\sum_{r\in R}c_{rs}c_{rn}^{*}J/(\gamma_{\mathrm{a}r}-i\Delta)$ describe couplings between different stable modes, and $\hat{G}_{\mathrm{a}s}=\hat{F}_{\mathrm{a}s}-iJ\sum_{r\in R}c_{sr}\hat{w}_{\mathrm{b}r}$, and $\hat{G}_{\mathrm{b}s}=\hat{F}_{\mathrm{b}s}-iJ\sum_{r\in R}c_{rs}\hat{w}_{\mathrm{a}r}$ denote the increased quantum noise induced due to coupling with the high-order (reservoir) modes $\mathcal{R}$.

The effect of $\hat{G}_{\mathrm{a}s}$ and $\hat{G}_{\mathrm{b}s}$ on the spin dynamics depends on the number of modes considered in $\mathcal{S}$.
For the case of an uncoated spherical cell with radius $R$, we can bound the contribution of the high-order modes by $|\epsilon_{sn}^{(\mathrm{a|b})}|<J/(\pi^{2}\gamma_{\mathrm{(b|a)}r_{0}})$.
Here $\gamma_{\mathrm{(b|a)}r_{0}}=D_{\mathrm{(b|a)}}\pi^{2}r_{0}^{2}/R^{2}$ is the diffusion-induced relaxation of the least decaying mode in the set $\mathcal{R}$, with a radial mode number $r_{0}$.
This bound is attained by the asymptotic form of the diffusion-relaxation modes, validated by numerical calculations.
Thus if enough modes are considered in Eqs.~(\ref{eq:stable-mode-a-EOM}), the contribution of $\epsilon^{(a)},\epsilon^{(b)}$, and $\hat{G}_{a}-\hat{F}_{a},\hat{G}_{b}-\hat{F}_{b}$ to the dynamics can be rendered negligible.
In general, this formalism can also be applied with a smaller number of stable modes such that $\gamma_{\mathrm{a}m},\gamma_{\mathrm{b}n}\sim J$, on the expense of overestimating the diffusion-induced relaxation.
Moreover, one may calculate $\epsilon^{(a)},\epsilon^{(b)}$ for a given number of leading reservoir modes and use the asymptotic approximation for the infinite number of additional reservoir modes, namely $\epsilon_{sn}^{(\mathrm{a})}=\sum_{r=r_{0}}^{r_{1}}c_{sr}c_{nr}^{*}J/(\gamma_{\mathrm{b}r}+i\Delta)+J/(\pi^{2}\gamma_{\mathrm{b}r_{1}})$ (and $\epsilon_{sn}^{(\mathrm{b})}=\sum_{r=r_{l}}^{r_{h}-1}Jc_{rs}c_{rn}^{\ast}/(\gamma_{\mathrm{a}r}-i\Delta)+J/\pi^{2}\gamma_{\mathrm{a}r_{h}}$).

Finally, we consider the analytical solution of Eqs.~(\ref{eq:coherent-dynamics}) in the main text corresponding to the simplifying case of an approximated two-mode solution, using a single stable mode $s=0$ (with $c_{00}=1$).
In the strong coupling regime $J\gg\gamma\equiv\gamma_{\mathrm{a}0}$, the full solution in the rotating frame reads
\begin{equation}
\begin{aligned}\hat{a}(t) & =e^{-\gamma t/2}[\cos(Jt)\hat{a}(0)-i\sin(Jt)\hat{b}(0)]+\hat{w}_{\mathrm{a}},\\
\hat{b}(t) & =e^{-\gamma t/2}[\cos(Jt)\hat{b}(0)-i\sin(Jt)\hat{a}(0)]+\hat{w}_{\mathrm{b}}.
\end{aligned}
\label{eq:stable-mode-decay-factor-2}
\end{equation}
Here the alkali-metal relaxation is shared by both spin gases, accompanied with a transfer of quantum fluctuations to the noble-gas spin ensemble.
The noise processes are defined by $\hat{w}_{\mathrm{a}}(t)=\int_{0}^{t}dse^{-\gamma s/2}\cos(Js)\hat{F}_{\mathrm{a}}(t-s)$ and $\hat{w}_{\mathrm{b}}(t)=-i\int_{0}^{t}dse^{-\gamma s/2}\sin(Js)\hat{F}_{\mathrm{a}}(t-s)$
using standard stochastic integration.
For $t\gg2/\gamma$, these processes can be considered as white Wiener operators, while, for shorter times, they are colored by the window functions.

\subsection{Collisions statistics}

Here we present details about the collisions statistics of alkali and noble-gas atoms, used in the derivation of Eq.~(\ref{eq:coarse-graind_equation_dJ_dt}) from Eq.~(\ref{eq:spin-operators-discrete-step-evolution1}) in the main text, as well as for characterization of the spin-exchange noise in Eq.~(\ref{eq:noise-operator}).

The microscopic parameter $\varkappa_{ab}\left(t,\tau\right)$ indicates if the pair $a-b$ has collided during the time interval $\left[t,t+\tau\right]$ by
\begin{equation}
\varkappa_{ab}\left(t,\tau\right)=\int_{t}^{t+\tau}\delta(s-t_{ab}^{\left(i\right)})ds.
\end{equation}
For short $\tau$, we can assume a ballistic motion of the particles, such that the two-body displacement satisfies 
\begin{equation}
\mathbf{r}_{ab}\left(t+s\right)=\mathbf{r}_{ab}\left(t\right)-\mathbf{v}_{ab}(t)s
\end{equation}
for any $s\in\left[t,t+\tau\right]$.
A collision of the pair occurs at $t_{ab}^{\left(i\right)}$ if $r_{ab}(t_{ab}^{\left(i\right)})\le\epsilon$, where $\epsilon$ characterizes the hard-sphere radius of the pair, satisfying $\sigma=\pi\epsilon^{2}$ and $r_{ab}=|\mathbf{r}_{ab}|$.
The time of a collision is then determined by
\begin{align*}
|\mathbf{r}_{ab}\left(t\right)-\mathbf{v}_{ab}\cdot(t_{ab}^{\left(i\right)}-t)| & \leq\epsilon.
\end{align*}
Solving for $t_{ab}^{(i)}$, we obtain the expression 

\[
t_{ab}^{\left(i\right)}=t+\frac{r_{ab}\left(t\right)}{v_{ab}}\left[\cos(\theta_{ab}^{v})\pm\sqrt{\epsilon^{2}/r_{ab}^{2}\left(t\right)-\sin^{2}(\theta_{ab}^{v})}\right],
\]
where $\theta_{ab}^{v}\in[0,\pi]$ is the relative angle between $\mathbf{r}_{ab}$ and $\mathbf{v}_{ab}$.
Therefore $t_{ab}$ exists only if $\sin^{2}(\theta_{ab}^{v})\leq\epsilon^{2}/r_{ab}^{2}\left(t\right)$.
Since $\epsilon$ is about a few angstroms, collisions occur only at small angles $\theta_{ab}^{v}\ll1$.
Neglecting the collision duration $\tau_\text{c}\apprle2\epsilon/v_{ab}\ll\tau$, we can approximate the collision time as the average of the two solutions, yielding $t_{ab}^{\left(i\right)}=t+r_{ab}\left(t\right)/v_{ab},$ if $\theta_{ab}^{v}\leq\epsilon/r_{ab}\left(t\right)$.
We can then write the expression of $\varkappa_{ab}(t,\tau)$ as 
\begin{equation}
\varkappa_{ab}\left(t,\tau\right)=\Theta\left(\theta_{ab}^{v}\leq\epsilon/r_{ab}\left(t\right)\right)\int_{0}^{\tau}\delta\left(s-r_{ab}\left(t\right)/v_{ab}\right)ds,
\end{equation}
which determines if a pair has collided given the relative location and velocities.

To derive the statistical properties of $\varkappa_{ab}$, we first average over the pairs velocities.
We assume a Maxwell-Boltzmann distribution for the velocity $\boldsymbol{v}$
\begin{equation}
f(\boldsymbol{v})d^{3}\boldsymbol{v}=\pi^{-\frac{3}{2}}\frac{v^{2}}{v_{\mathrm{T}}^{3}}e^{-v^{2}/v_{\mathrm{T}}^{2}}dv\sin\theta_{v}d\theta_{v}d\varphi_{v},
\end{equation}
where $v_{\mathrm{T}}$ stands for the thermal relative velocity of the pair.
The velocity-average collision probability is then given by
\begin{align*}
\langle\varkappa_{ab}\left(t,\tau\right)\rangle_{v} & \equiv\int\varkappa_{ab}\left(t,\tau\right)f(\boldsymbol{v})d^{3}\boldsymbol{v}\\
= & (\pi^{-\frac{3}{2}}/v_{\mathrm{T}}^{3})\int_{0}^{\infty}v^{2}e^{-v^{2}/v_{\mathrm{T}}^{2}}dv\int_{0}^{\epsilon/r_{ab}\left(t\right)}\sin\theta_{v}d\theta_{v}\int_{0}^{2\pi}d\varphi_{v}\int_{0}^{\tau}\delta\left(s-r_{ab}\left(t\right)/v\right)ds\\
= & \frac{\epsilon^{2}}{\sqrt{\pi}r_{ab}^{2}\left(t\right)v_{\mathrm{T}}^{3}}\int_{0}^{\infty}v^{2}e^{-v^{2}/v_{\mathrm{T}}^{2}}dv\int_{0}^{\tau}\delta\left(s-r_{ab}\left(t\right)/v\right)ds\\
= & \frac{\epsilon^{2}}{\sqrt{\pi}r_{ab}^{2}\left(t\right)v_{\mathrm{T}}^{3}}\int_{0}^{\tau}\frac{ds}{s}\int_{0}^{\infty}dvv^{3}e^{-v^{2}/v_{\mathrm{T}}^{2}}\delta\left(v-r_{ab}\left(t\right)/s\right)\\
= & \frac{\epsilon^{2}}{\sqrt{\pi}r_{ab}^{2}\left(t\right)v_{\mathrm{T}}^{3}}\int_{0}^{\tau}\frac{ds}{s^{4}}r_{ab}^{3}\left(t\right)e^{-r_{ab}^{2}\left(t\right)/(s^{2}v_{\mathrm{T}}^{2})}=\frac{\epsilon^{2}}{r_{ab}^{2}\left(t\right)}\frac{1}{\sqrt{\pi}}\int_{r_{ab}/(\tau v_{\mathrm{T}})}^{\infty}duu^{2}e^{-u^{2}},
\end{align*}
where in the last step we changed the integration variable to $u=r_{ab}\left(t\right)/(sv_{\mathrm{T}})$.
The last integral can be approximated using the Heaviside function 

\begin{equation}
\frac{1}{\sqrt{\pi}}\int_{r_{ab}/(\tau v_{\mathrm{T}})}^{\infty}duu^{2}e^{-u^{2}}\approx\frac{1}{4}\Theta(\tau v_{\mathrm{T}}-r_{ab}\left(t\right)),
\end{equation}
yielding

\begin{equation}
\langle\varkappa_{ab}\left(t,\tau\right)\rangle_{v}\approx\frac{\sigma}{4\pi r_{ab}^{2}\left(t\right)}\Theta(\tau v_{\mathrm{T}}-r_{ab}\left(t\right)),
\end{equation}
such that two particles collide, on average, depending on their relative solid angle $\sigma/(4\pi r_{ab}^{2}\left(t\right))$, provided that their spatial separation is small $r_{ab}<\tau v_{\mathrm{T}}$.
Our model relies on the motion of the particles being ballistic, which is valid for short intervals $v_{\mathrm{T}}\tau\ll1/(n_{\mathrm{b}}\sigma)$. 

We are interested in the spatially coarse-grained dynamics on the length-scale $l\gg1/(n_{\mathrm{b}}\sigma)$.
Using the radial window
function $w(\mathbf{r})$, we obtain

\begin{align}
\langle\varkappa_{ab}\left(t,\tau\right)\rangle_{v}*w(\mathbf{r}) & =\frac{3}{4\pi l^{3}}\int_{0}^{2\pi}d\phi'\int_{0}^{\pi}\sin\theta'd\theta'\int_{0}^{\infty}r'^{2}dr'\frac{\sigma}{4\pi r'^{2}}\Theta(\tau v_{\mathrm{T}}-r')\Theta(\left|\mathbf{r}_{ab}-\mathbf{r}'\right|-l)\nonumber \\
 & \approx\frac{3}{4\pi l^{3}}\sigma\tau v_{\mathrm{T}}\int_{0}^{2\pi}d\phi'\int_{0}^{\pi}\sin\theta'd\theta'\int_{0}^{\infty}r'^{2}dr'\frac{\delta(r')}{4\pi r'^{2}}\Theta(\left|\mathbf{r}_{ab}-\mathbf{r}'\right|-l)=\frac{3}{4\pi l^{3}}\sigma\tau v_{\mathrm{T}}\Theta(r_{ab}-l)\nonumber \\
 & =\sigma\tau v_{\mathrm{T}}w(r_{ab}),\label{eq:methods1}
\end{align}
where in the second line we used $v_{\mathrm{T}}\tau\lll l$.
This expression can be used to estimate standard kinematic relations, such as the mean collision times.
The probability that two spins $a-b$ would collide in time interval $\tau$ is given by $p_{ab}(t,\tau)=\langle\varkappa_{ab}\left(t,\tau\right)\rangle_{v}*w(\mathbf{r}).$
We can now find the mean time that particle $a$ has collided with any other noble gas atom.
This probability is given by
\[
p_{a}(t,\tau)=\sum_{b}p_{ab}(t,\tau)=\tau n_{\mathrm{b}}\sigma v
\]
using the relation $n_{\mathrm{b}}=\sum_{b}w(\mathbf{r}-\mathbf{r}_{b})$.
Since $\tau_{\mathrm{d}}^{(\mathrm{b})}\equiv1/(n_{\mathrm{b}}\sigma v)$ is the mean time between collision for a given alkali-metal atom with some other noble spin, the probability is simply $p_{a}=\tau/\tau_{\mathrm{d}}^{(\mathrm{b})}$, independent of $t$.
This result corresponds to the Markovian exponential distribution $p_{a}(t,\tau)=1-\exp(-\tau/\tau_{\mathrm{d}}^{(\mathrm{b})})$ for $\tau\ll\tau_{\mathrm{d}}^{(\mathrm{b})}$.
A similar result is obtained for $p_{b}(t,\tau)$ by interchanging the indices $a\leftrightarrow b$.

We calculate the second moment of $\varkappa_{ab}$ assuming that different collisions are statistically independent,

\begin{align}
\bigl\langle\varkappa_{ab}(t,\tau)\varkappa_{cd}(t',\tau)\bigr\rangle_{\text{v}} & =\delta_{ac}\delta_{bd}\Theta\left(\tau-\left(t-t'\right)\right)\bigl\langle\varkappa_{ab}(t,\tau)\varkappa_{ab}(t,\tau)\bigr\rangle_{\text{v}}\nonumber \\
= & \delta_{ac}\delta_{bd}\cdot\tau\delta\left(t-t'\right)\bigl\langle\left[P\left(\varkappa_{ab}(t,\tau)=1\right)\cdot1^{2}\right]\bigr\rangle_{\text{v}}\nonumber \\
= & \delta_{ac}\delta_{bd}\cdot\tau\delta\left(t-t'\right)\bigl\langle\varkappa_{ab}(t,\tau)\bigr\rangle_{\text{v}},\label{eq:methods2}
\end{align}
where in the second line we assumed that the times $t,t'$ are sampled with intervals $dt,dt'\gg\tau$ to include multiple collisions.

Spin exchange interactions, experienced during binary collisions as considered so far, lead to phase accumulation of the colliding spins.
The spin dynamics are determined by the statistics of the collisions and are governed by
\begin{equation}
\varkappa_{ab}\left(t,\tau\right)\phi_{ab}\left(t\right)=\int_{t}^{t+\tau}\phi_{ab}^{\left(i\right)}\delta(s-t_{ab}^{\left(i\right)})ds.
\end{equation}
Since $\varkappa_{ab}\left(t,\tau\right)$ is a Bernoulli process, whose possible values are only 0 or 1,we get 
\begin{align*}
\left\langle \varkappa_{ab}\left(t,\tau\right)\phi_{ab}\left(t\right)\right\rangle  & =P\left(\varkappa_{ab}\left(t,\tau\right)=1\right)\cdot1\cdot\left\langle \phi_{ab}\left(t\right)|\varkappa_{ab}\left(t,\tau\right)=1\right\rangle +P\left(\varkappa_{ab}\left(t,\tau\right)=0\right)\cdot0\cdot\left\langle \phi_{ab}\left(t\right)|\varkappa_{ab}\left(t,\tau\right)=0\right\rangle \\
 & =P\left(\varkappa_{ab}\left(t,\tau\right)=1\right)\left\langle \phi_{ab}\left(t\right)|\varkappa_{ab}\left(t,\tau\right)=1\right\rangle \\
 & =\left\langle \varkappa_{ab}\left(t,\tau\right)\right\rangle \left\langle \phi_{ab}\left(t\right)|\varkappa_{ab}\left(t,\tau\right)=1\right\rangle .
\end{align*}
Moreover, $\varkappa_{ab}\left(t,\tau\right)$ is defined such that when the spins $a,b$ do not approach each other closer then the collision distance $\epsilon$, then $\phi_{ab}\left(t\right)=0$, and $\phi_{ab}\left(t\right)\neq0$ only when $\varkappa_{ab}\left(t,\tau\right)=1$.
Therefore $\left\langle \phi_{ab}\left(t\right)|\varkappa_{ab}\left(t,\tau\right)=1\right\rangle =\langle\phi_{ab}\left(t\right)\rangle=\langle\phi\rangle,$ where $\langle\phi\rangle$ is averaged over all impact velocities and impact parameters.
The value of $\langle\phi\rangle$ and the dependence of $\phi_{ab}\left(t\right)$ on the collision trajectory (velocity and impact parameter) were discussed in \cite{Walter1998HapperWalkerPhiTrajectory}.
The averaged coupling strength is
\begin{equation}
\left\langle \varkappa_{ab}\left(t,\tau\right)\phi_{ab}\left(t\right)\right\rangle =\langle\phi\rangle\sigma v_{\mathrm{T}}\tau w(r_{ab}).\label{eq:methods3}
\end{equation}
Similarly, the averaged dissipation rate is
\begin{equation}
\left\langle \varkappa_{ab}\left(t,\tau\right)\phi_{ab}^{2}\left(t\right)\right\rangle =\langle\phi^{2}\rangle\sigma v_{\mathrm{T}}\tau w(r_{ab}),\label{eq:methods4}
\end{equation}
and the fluctuation in second-order is
\begin{align*}
\left\langle \varkappa_{ab}\left(t,\tau\right)\phi_{ab}\left(t\right)\varkappa_{cd}\left(t',\tau\right)\phi_{cd}\left(t'\right)\right\rangle  & =P\left(\varkappa_{ab}\left(t,\tau\right)=1,\varkappa_{cd}\left(t',\tau\right)=1\right)\left\langle \phi_{ab}\left(t\right)\phi_{cd}\left(t'\right)|\varkappa_{ab}\left(t,\tau\right)=1,\varkappa_{cd}\left(t',\tau\right)=1\right\rangle \\
 & =\left\langle \varkappa_{ab}\left(t,\tau\right)\varkappa_{cd}\left(t',\tau\right)\right\rangle \langle\phi_{ab}\left(t\right)\phi_{cd}\left(t'\right)\rangle\\
 & =\delta_{ac}\delta_{bd}\cdot\tau\delta\left(t-t'\right)\bigl\langle\varkappa_{ab}(t,\tau)\bigr\rangle\langle\phi_{ab}\left(t\right)\phi_{cd}\left(t'\right)\rangle\\
 & =\delta_{ac}\delta_{bd}\tau\delta\left(t-t'\right)\cdot\sigma v_{\mathrm{T}}\tau w(r_{ab})\cdot\langle\phi_{ab}^{2}\left(t\right)\rangle\\
 & =\delta_{ac}\delta_{bd}\cdot\sigma v_{\mathrm{T}}w(r_{ab})\langle\phi^{2}\rangle\tau^{2}\delta\left(t-t'\right).
\end{align*}
Eqs.~(\ref{eq:methods1}) and (\ref{eq:methods3}) are used to derive the properties of $\zeta$ in Eq.~(\ref{eq:coarse-graind_equation_dJ_dt}) in the methods section, and in the identity of the spin-exchange noise $\bigl\langle\boldsymbol{\hat{F}}_{\mathrm{ex}}(\mathbf{r},t)\bigr\rangle=0$.
Eqs.~(\ref{eq:methods2}) and (\ref{eq:methods4}) are used in deriving the properties of the incoherent exchange terms in Eq.~(\ref{eq:coarse-graind_equation_dJ_dt}) in the methods section and in the derivation of the variance identities of the spin-exchange noise in Eq.~(\ref{eq:noise-operator}).

\end{document}